\documentclass[prd,nofootinbib,superscriptaddress,preprintnumbers,balancelastpage,nofootinbib]{revtex4-2}

\usepackage{graphicx}  
\usepackage{subcaption}
\usepackage{dcolumn}   
\usepackage{bm}        
\usepackage{amssymb,amsmath}
\usepackage{slashed}   
\usepackage[colorlinks=true,allcolors=black]{hyperref}
\usepackage{multirow}

\usepackage{hyperref}
\usepackage{fontawesome5}

\usepackage{dsfont}
\usepackage{tabularx}
\usepackage[usenames,dvipsnames,svgnames,table]{xcolor}

\def\be{\begin{equation}}
\def\ee{\end{equation}}
\def\bea{\begin{eqnarray}}
\def\eea{\end{eqnarray}}

\newcommand{\Gcode}{$\texttt{Beyond21}$}

\newcommand{\Lya}{Ly-$\alpha$ }

\newcommand{\eps}{\epsilon}
\renewcommand{\vec}{\mathbf}

\definecolor{cborange}{HTML}{e69f00}
\definecolor{cbgreen}{HTML}{009e73}
\definecolor{cbyellow}{HTML}{f1dd42}
\definecolor{cblblue}{HTML}{56b4e9}
\definecolor{cbblue}{HTML}{0000FF}
\definecolor{defgrey}{HTML}{808080}
\definecolor{defgreen}{HTML}{008000}
\definecolor{defred}{HTML}{FA5F55}

\def\beq{\begin{equation}}
\def\eeq{\end{equation}}

\def\be{\begin{eqnarray}}
\def\ee{\end{eqnarray}}

\begin{document}
\widetext

\title{Beyond21: A Global Framework for Cosmic Dawn and Reionization Within and Beyond the Standard Model}

\author{Omer Zvi Katz}
\affiliation{Raymond and Beverly Sackler School of Physics and Astronomy, Tel-Aviv 69978, Israel}

\begin{abstract} 
Observations of the Cosmic Dawn (CD) and Epoch of Reionization (EoR) are steadily improving, opening new opportunities to study early galaxies through complementary probes.
To enable consistent interpretation of these observations, we present Beyond21, a fully open-source Python package that implements flexible prescriptions for Pop II and Pop III star formation and computes the resulting radiation backgrounds and their impact on the intergalactic medium. From this coupled evolution, Beyond21 predicts the global 21-cm signal, UV luminosity functions (UVLFs), the ionization history, and the contribution to the observed cosmic X-ray background (CXB) within a single, self-consistent pipeline. A full global evolution run executes in $\sim0.1 \ {\rm s}$ on a single CPU core, enabling broad, high-resolution parameter exploration. The modular architecture facilitates straightforward modification of astrophysical prescriptions and the incorporation of new physics. As an illustrative example, we implement a scenario in which a small fraction of dark matter is millicharged, leading to baryon cooling through elastic interactions.
\href{https://github.com/OmerZviKatz/Beyond21}{%
  \textcolor{black}{\faGithub}\ \textcolor{blue}{Beyond21}%
}
\end{abstract}

\maketitle
\noindent
\section{Introduction}
 
The emergence of the first stars and galaxies during the Cosmic Dawn (CD) initiated a profound transformation in the early universe, driving its heating, ionization, and chemical enrichment.
Although we have yet to directly observe these first luminous sources, their cumulative impact is imprinted on a broad and growing set of cosmological observables~\cite{dhandha2025exploitingsynergiesjwstcosmic, katz2025closingpopiiistarsconstraints, pochinda2024constrainingpropertiespopulationiii, Park:2018ljd,Fialkov:2016zyq,qin2020tale,lazare2024heraboundxrayluminosity}. 

Arguably one of the most promising probes of this era is the redshifted 21-cm signal, which traces the spin state of neutral hydrogen in the intergalactic medium (IGM). Through its strong dependence on the ionization, excitation, and thermal state of the IGM, the 21-cm signal is highly sensitive to the formation properties of early astrophysical sources and to their radiation spectra across a wide range of energies~\cite{dhandha2025exploitingsynergiesjwstcosmic, katz2025closingpopiiistarsconstraints, pochinda2024constrainingpropertiespopulationiii, pochinda2024constrainingpropertiespopulationiii, Park:2018ljd,Fialkov:2016zyq,qin2020tale,lazare2024heraboundxrayluminosity}. Although the current status of a CD global 21-cm detection—previously claimed by EDGES~\cite{Bowman_2018}—remains uncertain~\cite{Singh:2021mxo,Hills:2018vyr,Sims_2019,Murray_2022}, in the absence of a confirmed signal, null results from global 21-cm experiments could place meaningful constraints on astrophysical models, with the most recent measurement provided by the SARAS collaboration~\cite{Singh:2021mxo}. 

Other key observables include high-redshift galaxy surveys, in particular measurements of the evolving ultraviolet luminosity functions (UVLFs) obtained with the Hubble Space Telescope (HST)~\cite{Bouwens:2014fua,McLeod_2015,Oesch_2018,Bouwens_2021,Bouwens_2021} and, more recently, the James Webb Space Telescope (JWST)~\cite{donnan2024jwstprimernewmultifield,mcleod2023galaxyuvluminosityfunction,adams2024epochspaperiiultraviolet,robertson2024earliestgalaxiesjadesorigins,Harikane_2023}, which trace the underlying star formation activity~\cite{Madau:2014bja,Bouwens_2021,Park:2018ljd}. Additionally, although it does not provide explicit time dependent information, the unresolved component of the present-day cosmic X-ray background (CXB), currently measured by Chandra~\cite{Hickox_2006,Lehmer:2012ak,Cappelluti:2012rd,Harrison_2016}, sets an upper bound on the cumulative hard X-ray emission of early sources over cosmic time~\cite{Fialkov:2016zyq,pochinda2024constrainingpropertiespopulationiii,Katz:2024ayw,katz2025closingpopiiistarsconstraints}. Similarly, reionization observables such as the CMB optical depth to reionization, $\tau_{\rm e}$, inferred by Planck~\cite{Planck:2018vyg}, and absorption lines in quasar spectra~\cite{McGreer:2014qwa,Davies_2025} encode valuable information about the ionizing properties of early astrophysical sources. 

While the CD observables discussed above are primarily used to study early astrophysical sources, they are also sensitive to scenarios involving new physics. Such scenarios may manifest through additional contributions to radiation fields, for example from dark matter annihilation or decay~\cite{Chen_2004, Agius:2025nfz,Facchinetti_2024,cima2025probingnonminimaldarksectors,Evoli_2014,Lopez_Honorez_2016,Liu_2018,Liu_2020}; through direct interactions with the intergalactic medium, such as baryon–dark matter scattering~~\cite{Tashiro:2014tsa,Munoz:2015bca,Munoz:2018pzp,Barkana:2018lgd,Barkana:2018qrx,Driskell:2022pax,Berlin:2018sjs,Liu:2019knx,Barkana:2022hko,Flitter_2024}; or via modifications to structure formation and star-formation processes, including suppressed small-scale power or delayed halo collapse in models such as fuzzy dark matter or warm dark matter~\cite{Hu_2000,villanuevadomingo2021sheddinglightdarkmatter,Flitter_2022}. For a wider discussion also see e.g.~\cite{Katz:2024ayw}. 

Relating this diverse set of observables to underlying models (either standard or new physics) requires tracing the coupled evolution of stellar populations, radiation backgrounds, and the IGM throughout cosmic history. This task has been addressed by a number of codes in the literature~\cite{Mesinger:2010ne,Visbal_2012,Mu_oz_2023,Flitter_2024,sun2023inhomogeneousenergyinjection21cm}, relying on simplified modeling and semi-analytical approaches. Here, we introduce a new global evolution code, \Gcode, which goes beyond the 21-cm output and combines predictions for the global 21-cm signal, ionization history, UV luminosity functions, and X-ray background within a single, self-consistent pipeline.

In addition to enabling joint predictions for multiple observables, \Gcode\ provides several advantages for both modeling and inference studies. Its short runtime, $\sim0.1 \ \mathrm{s}$ on a single CPU core, enables high dimensional inference analyses with minimal computational resources.
The simple and modular architecture, described in App.~\ref{Sec:Structure}, further allows for the straightforward implementation of new astrophysical models or extensions beyond the Standard Model (BSM) in a transparent and controlled manner. Moreover, \Gcode\ can be initialized at a very high redshift, up to $z\sim1800$ providing a single pipeline that evolves the IGM from before hydrogen recombination through the end of reionization at $z\sim6$. This feature is typically not required for standard astrophysical studies, but is often essential when considering new physics.

The structure of this paper is as follows. In Sec.~\ref{sec:observables}, we provide an overview of the main global observable outputs in \Gcode. Section~\ref{Sec:AstroModel} then describes the default astrophysical modeling currently implemented in \Gcode \  and the resulting radiation fields. In Sec~\ref{Sec:IGMEvolve}, we present the evolution of the IGM and its response to radiation. The following section~\ref{Sec:BSM} demonstrates a BSM implementation, where we incorporate the two coupled dark matter (2cDM) model of~\cite{Liu:2019knx} and examine the interplay between its 21-cm signatures and the astrophysical framework. We conclude in Sec.~\ref{sec:Summary}.

\section{Primary Outputs -  Observables}\label{sec:observables}

The main goal of \Gcode \ is to relate astrophysical and, if desired, also new physics models to observables from CD 
by tracing the global evolution of stellar populations and the intergalactic medium across cosmic time. To provide context for the intermediate quantities computed within \Gcode{}, we first review its main observable outputs.

\subsection{The Global 21-cm Signal}\label{sub:T21}

As CMB photons traverse the predominantly neutral IGM—from the Dark Ages to the EoR—the hyperfine transitions of ground-state hydrogen, accompanied by 21-cm photon absorption and emission, imprint a signal over the background spectrum. This signal is described by the 21-cm brightness temperature (for reviews see~\cite{Pritchard:2011xb,10.1088/2514-3433/ab4a73,Liu:2022iyy,villanuevadomingo2021sheddinglightdarkmatter})
\begin{align}\label{eq:T21}
    T_{21}(z) &= \frac{T_s(z)-T_\gamma(z)}{1+z}\left(1-e^{-\tau_{21}}\right) \,,
\end{align}
where~\cite{10.1088/2514-3433/ab4a73,villanuevadomingo2021sheddinglightdarkmatter} 
\begin{equation}\label{eq:21depth}
    \tau_{21}(z) = \frac{3}{32\pi}\frac{hc^3A_{10}}{k_{\rm B}T_{\rm s}(z)\nu_{10}^2}\frac{x_{\rm HI}(z)n_{\rm H}(z)}{(1+z)H(z)}
\end{equation}
is the optical depth of 21-cm photons from redshift $z$ to present day, with $A_{10}=2.85\times10^{-15}\ {\rm s}^{-1}$ the Einstein sponteneous emission coefficient of the hyperfine transition, $T_\gamma=2.75(1+z)$K the background temperature, and $x_{\rm HI}(z) \equiv n_{\rm HI}/n_{\rm H}$ the fraction of hydrogen that is neutral. The spin (effective) temperature $T_{\rm s}$ is defined through the relative occupation of the hyperfine triplet and singlet states of hydrogen, $n_1/n_0\equiv3e^{-h\nu_{10}/T_{\rm s}}$, where $\nu_{10}$ is the energy splitting between the states, corresponding to 21-cm.

In standard cosmology, the spin temperature is balanced between three spin-flipping processes that control the population of singlet and triplet hydrogen states (see e.g.~\cite{Furlanetto:2006jb,Pritchard:2011xb}):
\begin{itemize}

    \item At redshifts \( z \gtrsim 100 \), spin-flip transitions are dominated by hydrogen collisions with the different IGM components. At these redshifts, the collisional excitation and de-excitation rates greatly exceed the Hubble expansion rate, driving the hyperfine level populations into thermal equilibrium with the kinetic temperature of the hydrogen gas, \( T_{\rm k} \). As a result, the spin temperature \( T_{\rm s} \) approaches \( T_{\rm k} \). The collisional de-excitation rate is
    \begin{equation}
        \Gamma^{\rm col}_{1\rightarrow0} = \sum_i n_i \kappa_i(T_{\rm k}),
    \end{equation}
    where \( n_i \) is the number density of specie \( i = {\rm HI,\, e,\, p} \), neglecting helium, and the temperature dependent factors \( \kappa_i(T_{\rm k}) \) are provided in~\cite{Furlanetto:2006su,Zygelman_2005,Allison1969,Pritchard:2006sq}.

    \item As the IGM expands and cools, radiative 21-cm transitions induced by CMB photons eventually overtake collisional processes. These transitions occur faster than the Hubble rate, with an induced de-excitation rate
    \begin{equation}
        \Gamma^\gamma_{1\rightarrow0} =  B_{10} u_\nu(E_{21}) \approx A_{10} \frac{k_{\rm B}T_\gamma}{E_{21}} \,,
    \end{equation}
    thereby coupling the spin temperature to the CMB temperature \( T_\gamma \). Here, \( B_{10} \) is the stimulated de-excitation Einstein coefficient, and \( u_\nu(E_{21}) \) is the CMB spectral energy density at \( E_{21} \). The inverse process is controlled by $B_{01} = 3B_{10}$.

    \item Eventually, stars form, producing \Lya photons either through the redshifting of stellar UV radiation directly into the \Lya frequency or through excitations to higher Lyman levels which are then followed by radiative cascades~\cite{Hirata:2005mz} (see also Sec.~\ref{SubSec:Radiation} ). These \Lya photons can induce spin-flip transitions in neutral hydrogen: a hydrogen atom in one of the 1s hyperfine states absorbs a \Lya photon, then re-emits it and decays to the opposite hyperfine state with some probability—a process commonly referred to as the Wouthuysen–Field (WF) effect~\cite{1958PIRE...46..240F,1952AJ.....57R..31W,Hirata:2005mz}. It is customary to define the \Lya color temperature, \( T_\alpha \), as the effective temperature driven by these transitions, such that
    \[
    \frac{\Gamma_{0\rightarrow1}^{\alpha}}{\Gamma_{1\rightarrow0}^{\alpha}} = 3e^{-E_{21}/k_{\rm B}T_\alpha} \,.
    \]
    The \Lya-induced spin-flip transition rates between the hyperfine levels \( i \rightarrow j \) depend on the global photon flux around the \Lya frequency and are given by
    \begin{equation}
        \Gamma^\alpha_{i\rightarrow j} = 4\pi \int \sigma_{ij}(\nu)J(\nu) \, d\nu 
        \equiv \frac{32\pi}{9} \lambda_\alpha^2 \gamma_\alpha \tilde{S}_{ij} J_\alpha \,,
        \label{eq:LyaRate}
    \end{equation}
    where $\sigma_{ij}(\nu) = \frac{3}{2}\lambda^2_\alpha\gamma_\alpha\phi_{ij}(\nu) $ is the spin-flip cross section, $\lambda_\alpha$ is the wavelength of \Lya photons, $\gamma_\alpha=50\text{ MHz}$ is the line half-width at half-maximum, $\phi_{ij}(\nu)$ is the corresponding line profile, and $J(\nu)$ is the specific number intensity (number of photons per time, frequency, area and steradian) around the \Lya energy. The spectrum near the \Lya line center is altered by energy exchange between \Lya photons and hydrogen atoms due to recoil and spin flips in WF interactions~\cite{Chen:2003gc,Hirata:2005mz}. $J_\alpha = J_\alpha^\star + J_\alpha^{\rm X}$
    represents the smooth, astrophysically sourced \Lya{} background prior to the inclusion of resonant \Lya scattering effects. Its two contributions originate from UV and X-ray sources, and are modeled in Secs.~\ref{sec:Nonion_UV} and~\ref{subsubsec:Xrays}, respectively.

    Accounting for spectral distortions around the \Lya line requires solving a resonant-scattering Fokker–Planck equation~\cite{Hirata:2005mz} and is computationally expensive; to avoid repeating this calculation, for every set of astrophysical parameters, it is common to introduce an astrophysics-independent, dimensionless factor
    \[
    \tilde{S}_{ij} = \frac{27}{16} \int \phi_{ij}(\nu) \frac{J(\nu)}{J_\alpha} \, d\nu \,.
    \]
    \Gcode\ uses pre-computed interpolation functions for \( \tilde{S}_{ij}(T_{\rm k},T_{\rm s}, \tau_{\rm GP}) \) and \( T_\alpha (T_{\rm k},T_{\rm s}, \tau_{\rm GP})\) with $\tau_{\rm GP} = \frac{3 n_{\rm H} x_{\rm HI} \lambda_\alpha^3 \gamma_\alpha}{2 H}$ the Gunn–Peterson optical depth. The computation follows Ref.~\cite{Hirata:2005mz}\footnote{Ref.~\cite{Hirata:2005mz}  defines a total correction factor \( \tilde{S} = \tilde{S}_{10} + \tilde{S}_{01} \).}.
    
\end{itemize}

Solving the steady-state Boltzmann equation for the abundances of the hyperfine states, given the above rates, and assuming $E_{21} \ll T_\gamma, T_{\rm k}, T_\alpha$ we acquire the standard equation for the spin temperature evolution 
\begin{equation}\label{eq:Ts}
T_s^{-1}=\frac{ T_{\gamma}^{-1}+\bar x_\alpha T_{\alpha}^{-1} +\bar x_k T_{k}^{-1}}{1+\bar x_\alpha+\bar x_k}\,,
\end{equation}
where $\bar{x}_\alpha\equiv\frac{\Gamma^\alpha_{0\rightarrow1} + \Gamma^\alpha_{1\rightarrow0}}{\Gamma^\gamma_{0\rightarrow1} + \Gamma^\gamma_{1\rightarrow0}}$ and $\bar{x}_{\rm col}\equiv\frac{\Gamma^{\rm col}_{0\rightarrow1} + \Gamma^{\rm col}_{1\rightarrow0}}{\Gamma^\gamma_{0\rightarrow1} + \Gamma^\gamma_{1\rightarrow0}}$.

\begin{figure*}[h!t]
  \centering
  
  \hfill
  \begin{subfigure}[t]{0.48\textwidth}
  \includegraphics[width=\textwidth]{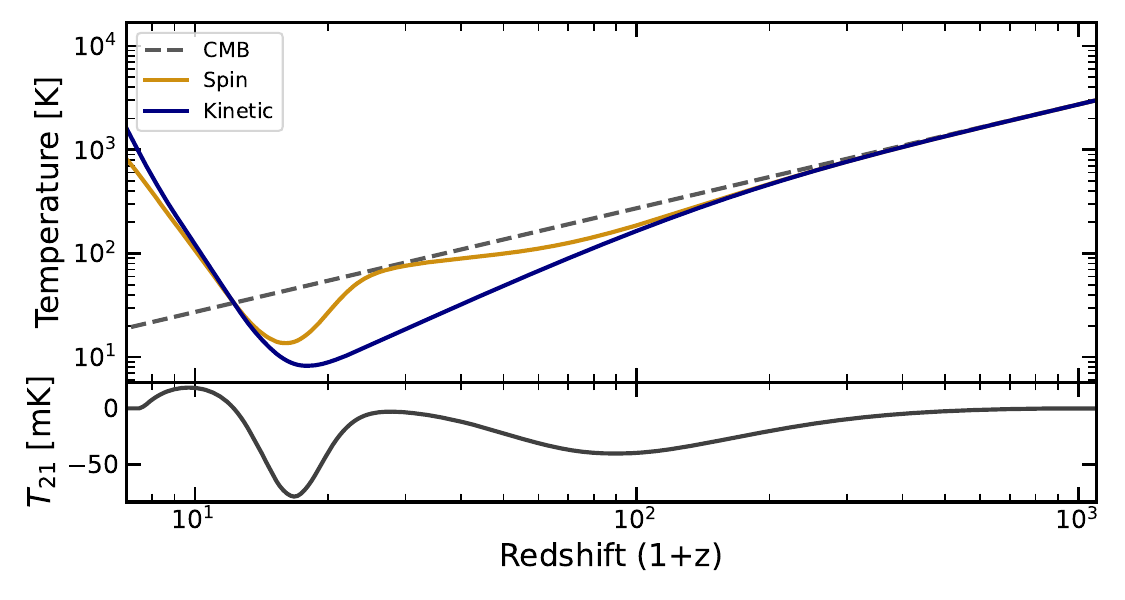}
  \end{subfigure}
  \begin{subfigure}[t]{0.48\textwidth}
    \includegraphics[width=\textwidth]{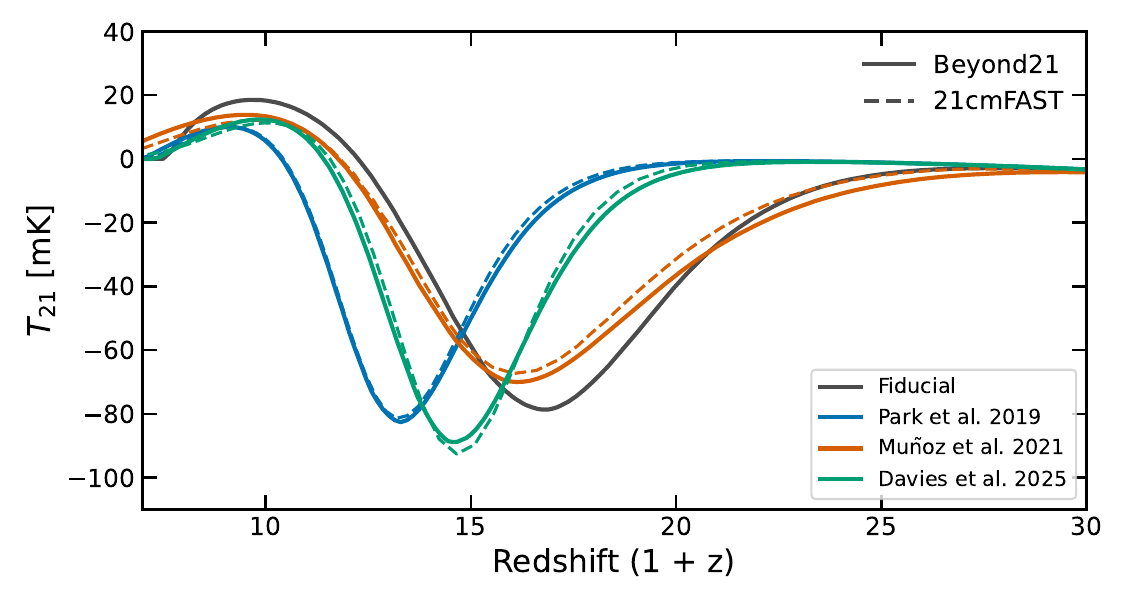}
  \end{subfigure}

  \caption{\textbf{Left (upper)}: Evolution of the spin (gold) and kinetic (navy) temperatures for our fiducial model (see Tab.~\ref{tab:Params}), together with the CMB temperature (dashed). \textbf{Left (lower)}: The resulting global 21-cm signal.
  At high redshifts, the CMB, kinetic, and spin temperatures are tightly coupled scaling as $T \propto (1+z)$,  yielding a null 21-cm signal (see Eq.~\eqref{eq:T21}). 
    Around $z \sim 250$, the kinetic temperature decouples from the CMB and evolves adiabatically, scaling as $T \propto (1+z)^{2}$ (see Sec.~\ref{SubSec:Tk}). The spin temperature remains coupled to the cooling gas through collisions, producing the first 21-cm absorption trough, which peaks near $z \sim 100$ when collisional coupling weakens and induced CMB interactions drive $T_{\rm s}$ back toward $T_{\rm CMB}$. 
    At $z \sim 30$, \Lya photons from the first stars couple $T_{\rm s}$ to the effective Ly$\alpha$ temperature, generating a second, deeper absorption feature in $T_{21}$. 
    Subsequently, X-ray heating increases the gas and \Lya temperatures, driving $T_{\rm s}$ above $T_{\rm CMB}$ and producing an emission signal. 
    Finally, as reionization progresses ($z \lesssim 10$), ionizing UV photons deplete neutral hydrogen, causing the 21-cm signal to vanish independently of $T_{\rm s}$ (Eq.\eqref{eq:T21}).
  \textbf{Right}: Comparison between \Gcode\ (solid) and 21cmFAST (dashed) global 21-cm evolutions for three benchmark models. Brown curves correspond to the fiducial two-population (Pop~II and Pop~III) model of \cite{Munoz:2021psm}, while blue lines follow the single-population fiducial model of \cite{Park:2018ljd}. The green curves show the evolution predicted by the latest 21cmFAST model~\cite{Davies_2025_1}, which assumes stochastic galactic properties and a continuous metallicity evolution. }
  \label{fig:T21}
\end{figure*}

In Fig.~\ref{fig:T21} (left) we show the redshift evolution of the spin temperature, kinetic temperature, and the resulting global 21-cm brightness temperature for our fiducial set of astrophysical and cosmological parameters (see Tab.~\ref{tab:Params}).
Figure~\ref{fig:T21} (right) compares the predictions of \Gcode \ with those of \textsc{21cmFAST} for three benchmark astrophysics models~\cite{Park:2018ljd,Munoz:2021psm, Davies_2025_1}, showing excellent agreement.

During the CD, the 21-cm signal is strongly dependent on astrophysical processes, rendering it a powerful diagnostic of early galaxy formation and evolution~\cite{dhandha2025exploitingsynergiesjwstcosmic, katz2025closingpopiiistarsconstraints, pochinda2024constrainingpropertiespopulationiii, pochinda2024constrainingpropertiespopulationiii, Park:2018ljd,Fialkov:2016zyq,qin2020tale,lazare2024heraboundxrayluminosity}. The coupling between the spin temperature and the color temperature, mediated by UV radiation from astrophysical sources results in absorption with respect to the CMB. The kinetic temperature of neutral hydrogen---on which \(T_\alpha\) depends, and to which it couples in the limit of infinite WF scattering events---is further regulated by X-ray emission that heats the IGM, affecting the depth and width of the absorption signal, often resulting in a consequent emission signal. Moreover, both UV and X-ray photons contribute to hydrogen ionization, progressively reducing the neutral fraction and thereby attenuating the 21-cm signal (see Eqs.~\eqref{eq:T21} and~\eqref{eq:21depth}). 
In Secs.~\ref{Sec:AstroModel} and~\ref{Sec:IGMEvolve} we describe the astrophysical models applied in \Gcode, and their impact on the IGM properties.

\subsection{The Cosmic X-ray Background}\label{sub:CXB}
Measurements of the CXB, currently by Chandra~\cite{Hickox_2006,Lehmer:2012ak} and expected by Athena~\cite{Marchesi:2020smf}, provide a valuable constraint on the residual X-ray intensity from high redshift sources~. A conservative upper limit is obtained by attributing the entire unresolved CXB, after subtraction of resolved low-redshift sources, to the modeled high-redshift emission~\cite{Fialkov:2016zyq,pochinda2024constrainingpropertiespopulationiii,Lehmer:2012ak,Cappelluti:2012rd, katz2025closingpopiiistarsconstraints,Katz:2024ayw}.

The \texttt{CXB} method in \Gcode\ computes the contribution of X-ray sources at redshifts $z>z_{\rm X}$ to the intensity measured today over a user-specified energy interval $[E_{\rm low}, E_{\rm high}]$,
\begin{equation}
\label{eq:CXB}
    I_{[E_{\rm low},\,E_{\rm high}]} = \frac{1}{\left(1+z_{\rm X}\right)^3} 
    \int_{\nu_{\rm low}}^{\nu_{\rm high}} I_{\rm X}\!\left(z_{\rm X}, \nu(1+z_{\rm X})\right) \,  e^{-\tau_{\rm X}(\nu,z_{\rm X})}\, d\nu \,,
\end{equation}
where $I_{\rm X}(z,\nu)$ is the global X-ray specific intensity (energy per unit time, frequency, area and solid angle) at redshift $z$ and rest frame frequency $\nu$, including the contributions from all sources at $z'\geq z$. \(\tau_{\rm X}(\nu,z_{\rm X})\) is the optical depth accumulated by X-rays observed today at frequency \(\nu\) along their propagation from redshift \(z_{\rm X}\).
This attenuation is dominated by photoionization, whose cross section increases
toward lower energies and therefore primarily affects the soft part of the spectrum~\cite{McQuinn_2012,Verner:1996th}.
Attenuation at \(z > z_{\rm X}\) is already accounted for in the intensity evaluated at $z_{\rm X}$.

For most scenarios of interest, one can safely neglect \( \tau_{\rm X}\) in Eq.~\eqref{eq:CXB}.
After reionization, absorption in the IGM is negligible, while prior to it (at
\( z \gtrsim 6 \)), even the photons observed today at energies as low as
\( E \sim 0.5\,\mathrm{keV} \) were sufficiently energetic so that
photoionization was strongly suppressed~\cite{McQuinn_2012,Verner:1996th}. Moreover, CXB constraints are often derived from high-latitude fields with low Galactic column densities~\cite{2016,Hickox_2006,Harrison_2016}, where also the attenuation in the Milky Way ISM is weak~\cite{Wilms_2000}.

Nevertheless, for completeness, \Gcode \ accounts for the full X-ray attenuation.
The IGM contribution is computed self-consistently from the internally evolved,
redshift-dependent \([\mathrm{HI},\mathrm{HeI},\mathrm{HeII}]\) abundances (see Sec.~\ref{subsubsec:Xrays}),
while absorption in the Milky Way ISM is modeled as
\begin{equation}
\tau_{\rm X}^{\rm MW}(\nu,N_{\rm H}) =
N_{\rm H}\left[\sum_{Z} A_Z\,\sigma_Z(\nu)\;+\;A_{\mathrm{H}_2}\,\sigma_{\mathrm{H}_2}(\nu)\right]\,,
\end{equation}
summing over the contributions of atomic species and molecular hydrogen. Depletion into dust grains is expected to have only a negligible effect on the optical depth~\cite{Wilms_2000} and is therefore omitted. Here, $A_i=N_i/N_{\rm H}$ denotes the abundance of species $i$ relative to hydrogen, and $\sigma_i(\nu)$ is the corresponding photoionization cross section. 
ISM abundances are taken from~\cite{Wilms_2000}, atomic photoionization cross sections from~\cite{verner1995analytic}, and the H2 cross section from~\cite{yan1998photoionization}. The total Milky Way hydrogen column density $N_{\rm H}$ is provided as an input to the \texttt{CXB} method.

In Sec.~\ref{Sec:AstroModel} we describe the default high-redshift X-ray source model implemented in \Gcode{} and derive the resulting global X-ray intensity $I_{\rm X}(z,\nu)$, thereby establishing an explicit connection between the model’s astrophysical parameters and its predicted CXB. This relation is explored in depth in an upcoming publication~\cite{Xray}.

\subsection{High Redshift UVLFs}\label{sub:UVLFs}

The \texttt{UVLF} module in \Gcode\ computes the rest-frame UV luminosity function at a given redshift, enabling direct comparison with HST~\cite{Bouwens:2014fua,Bouwens_2021}  and JWST~\cite{Harikane_2023,donnan2024jwstprimernewmultifield,robertson2024earliestgalaxiesjadesorigins} observations, shedding light on the abundance, star formation activity, and UV properties of early galaxies. Defined as the comoving number density of galaxies
per rest-frame UV magnitude $M_{\rm UV}$, and assuming at most one UV-emitting galaxy
per dark matter halo, the UVLF can be written as
\begin{equation}
    \phi(z,M_{\rm UV}) = \int dM_{\rm h}\frac{dn(M_{\rm h},z)}{dM_{\rm h}}f_{\rm gal}(M_{\rm h},z) P(M_{\rm UV}|M_{\rm h},z)\,,
    \label{eq:UVLF}
\end{equation}
where $M_{\rm h}$ denotes halo mass and $dn/dM_{\rm h}$ the halo mass function (HMF). 
The factor $f_{\rm gal}(M_{\rm h},z)$ is a smooth halo-mass filter that encodes the cooling requirements for star formation in low-mass halos, as discussed in Sec.~\ref{SubSec:SFR}, while 
$P(M_{\rm UV}\!\mid\!M_{\rm h},z)$ describes the distribution of UV magnitudes at fixed halo mass and redshift. In its default configuration, \Gcode\ 
adopts either a deterministic\footnote{For generality, we avoid the commonly used deterministic form
$
\phi(z,M_{\rm UV}) =
f_{\rm gal}(M_{\rm h},z)\,
\frac{dn(M_{\rm h},z)}{dM_{\rm h}}
\left|\frac{dM_{\rm h}}{dM_{\rm UV}}\right|,
$
which assumes a monotonic mapping $M_{\rm UV}(M_{\rm h})$ at fixed $z$.}  halo--magnitude mapping $M_{\rm UV}(M_{\rm h},z)$ or a Gaussian scatter with standard deviation $\sigma_{\rm UV}$ about this mean relation, corresponding to log-normal scatter in luminosity.

The mapping between halo mass and UV magnitude follows directly from the star-formation prescription introduced in Sec.~\ref{SubSec:SFR}. The star formation rate $\dot{M}_\star(M_{\rm h},z)$ is converted to a rest-frame UV luminosity at $\lambda \simeq 1500 \ \text{\AA}$ using the approximately linear calibration derived from stellar population synthesis models~\cite{Kennicutt_1998,Madau:2014bja}
\begin{equation}
    \dot{M}_\star(M_{\rm h},z) = \kappa_{\rm UV} L_{\rm UV} \,,
    \label{eq:MpropL}
\end{equation}
where $\kappa_{\rm UV}$ encodes stellar population properties such as IMF, age, and metallicity, and is user-specified in \Gcode, allowing different values for galaxies hosting different stellar populations\footnote{To maintain full consistency, the coefficients $\kappa_{\rm UV}$ should be adopted in accordance with the UV spectra assumed in Sec.~\ref{SubSec:Radiation}.}. The luminosity is converted to absolute AB magnitude by~\cite{oke1983secondary}
\begin{equation}
    \log_{\rm 10}\left(\frac{L_{\rm UV}}{\rm erg \ s^{-1}\ Hz^{-1}}\right) = 0.4\times(51.63-M_{\rm UV}) \,.
\end{equation}
Together, these relations define the mean halo–magnitude mapping $M_{\rm UV}(M_{\rm h},z)$.

Our current treatment of the UVLF does not include attenuation by dust. However, dust effects are expected to be modest at \( z \gtrsim 6 \) for low and intermediate luminosity galaxies, becoming increasingly important only toward the bright, massive end. Our results should therefore remain reliable for \( M_{\rm UV} \gtrsim -20 \) at \( z \gtrsim 6 \), and likely extend to even brighter magnitudes at higher redshifts~\cite{Zhao_2024,cullen2024ultravioletcontinuumslopeshighredshift,donnan2024jwstprimernewmultifield,dhandha2025exploitingsynergiesjwstcosmic}. A self-consistent treatment of dust attenuation will be incorporated in future extensions of \Gcode.

\begin{figure}[h!]
    \centering
    \includegraphics[width=0.65\textwidth]{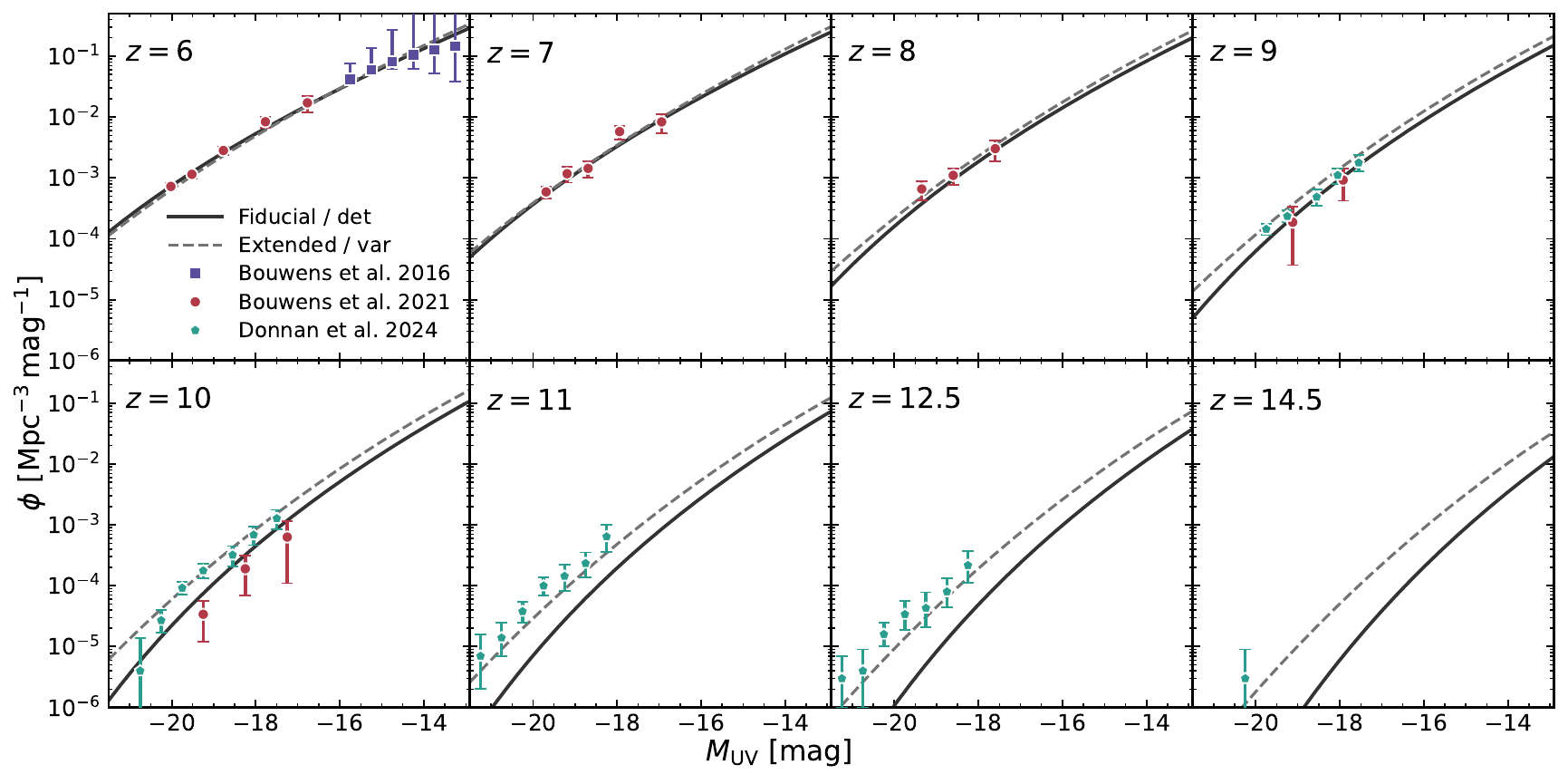}
    \caption{UVLFs produced with \Gcode. \textbf{Solid} curves correspond to the fiducial model, which assumes a deterministic halo–magnitude relation and is calibrated using the UVLFs observed by the HST for $M_{\rm UV} > -20\,\mathrm{mag}$ at $6 \leq z \leq 10$.
    \textbf{Dashed} curves show an extended model  that includes a Gaussian scatter about the mean halo–magnitude relation and is calibrated to the HST dataset at $z\leq8$ and to JWST measurements at higher redshifts. The model parameters are given in Tab.~\ref{tab:Params}.
    \textbf{Red} and \textbf{purple} data points denote HST measurements from~\cite{Bouwens:2014fua,Bouwens_2021}, while \textbf{turquoise} show JWST data from~\cite{donnan2024jwstprimernewmultifield}} 
    \label{fig:UVLF}
\end{figure}

In Fig.~\ref{fig:UVLF} we present the UVLFs predicted by our fiducial astrophysical model together with HST~\cite{Bouwens:2014fua,Bouwens_2021} and JWST~\cite{donnan2024jwstprimernewmultifield} data. To maintain a minimal and controlled setup, the model is calibrated to the UVLFs using HST data alone and does not attempt to fit current JWST measurements (see App.~\ref{Sec:Model}). It assumes a deterministic halo mass–UV magnitude relation and adopts the standard calibrated value of $\kappa_{1500 \text{\AA}} = 1.15\times10^{-28} \frac{M_\odot \ \text{yr}^{-1}}{\text{erg} \ \text{s}^{-1} \ \text{Hz}^{-1}}$ derived by~\cite{Madau:2014bja}, independently of redshift or stellar population.

Notably, JWST has reported a higher abundance of bright high-redshift galaxies than anticipated by many pre-JWST models, although the physical origin of this discrepancy remains unsettled. Proposed explanations range from redshift-dependent star-formation efficiency~\cite{Li_2024,kar2025extremeburstinessevolvingstar} and evolution in the UV–SFR conversion~\cite{Inayoshi_2022,Liu_2025} to variability in early galaxy properties~\cite{shen2023impactuvvariabilityabundance,kravtsov2024stochasticstarformationabundance,furlanetto2022bursty}. The modular design of \Gcode\ allows such modifications to be implemented straightforwardly without changing the overall framework. As a simple illustrative example, we also consider a model that includes a scatter in UV magnitude. Fitting to the combined HST and JWST data yields a best-fit value for a scatter of $\sigma_{\rm UV} = 1.7 \ {\rm mag}$. The resulting UVLFs are shown in Fig.~\ref{fig:UVLF}, with the corresponding model parameters listed in Tab.~\ref{tab:Params}.

\subsection{Reionization Observables}
\label{sub:ReionObs}

\begin{figure}[h!]
    \centering
    \includegraphics[width=0.4\textwidth]{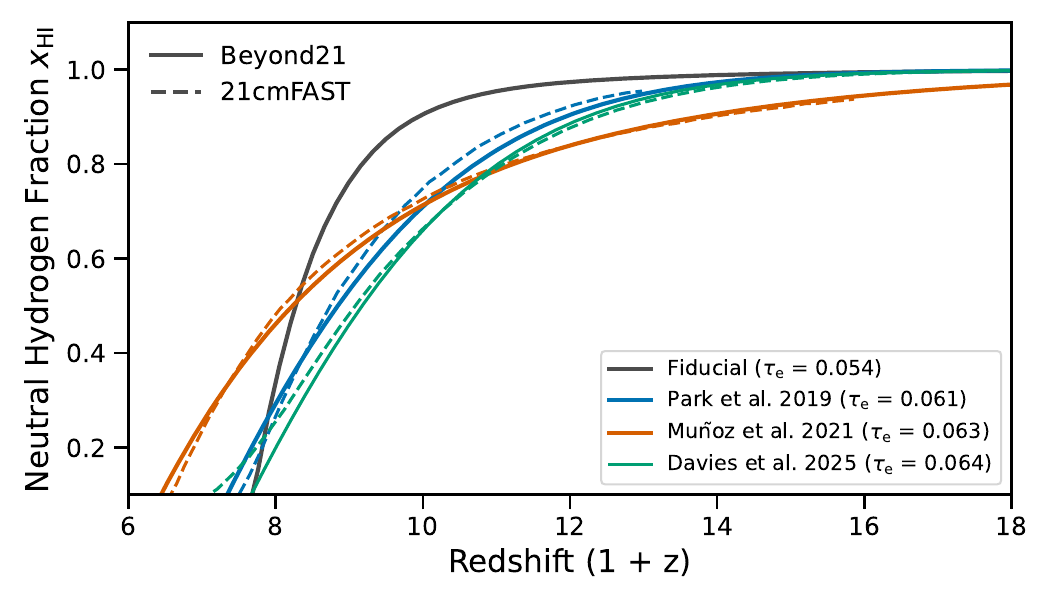}
    \caption{The evolution of the neutral hydrogen fraction. Results for our fiducial model (Tab~\ref{tab:Params}) is shown in \textbf{grey}. \textbf {Blue}~\cite{Park:2018ljd}, \textbf{orange}~\cite{Munoz:2021psm} and \textbf{green}~\cite{Davies_2025_1} are the $x_{\rm HI}$ evolutions for the same 21cmFAST models as in fig~\ref{fig:T21}. \textbf{Solid} are ionization histories generated with \Gcode, while \textbf{dashed} are outputs from 21cmFAST. The electron scattering optical depth of CMB photons for each model is given in the legend. } 
    \label{fig:xHI}
\end{figure}

\Gcode{} evolves and outputs the global ionization fraction of hydrogen and helium throughout cosmic history, as detailed in Sec.~\ref{Sec:IGMEvolve} and shown in Fig.~\ref{fig:xHI}. These quantities depend sensitively on the astrophysical model, since UV and X-ray emission from early sources drive the EoR.

For convenience, \Gcode \ computes the dominant model independent observable related to the ionization fraction, which is the optical depth of CMB photons
\begin{equation}
\tau_{e}=n_H(z=0) \sigma_T\int^{50}_{0} d z x_e(z)\frac{(1+z)^2}{H(z)}\, ,\label{eq:opticaldepth}
\end{equation}
where $\sigma_{\rm T}$ is the Thomson cross section and $x_{\rm e}=n_{\rm e}/n_{\rm H}$ the total electron fraction. 
This output can be directly compared to the \textit{Planck} measurement
\(\tau_{\rm e} = 0.054 \pm 0.007\) at 68\% C.L.~\cite{Planck:2018vyg}.

While the CMB optical depth only constrains the integrated ionization history, additional observables, such as absorption lines in quasar spectra~\cite{McGreer:2014qwa,Davies_2025}, which imply $
x_{\mathrm{HI}} \le 
\{\,0.030 + 0.048,\;
      0.095 + 0.037,\;
      0.191 + 0.056,\;
      0.199 + 0.087\,\}$ 
at redshifts
$z = \{\,5.481,\; 5.654,\; 5.831,\; 6.043\,\}$, may provide additional information on the ionized fraction at specific redshifts. For a review on reionization observables see~\cite{Mason_2019}.

\section{Astrophysical Modeling - Building a Global 
Stellar and radiative History}\label{Sec:AstroModel}

\Gcode \  traces the global radiation fields in both the UV and X-ray regimes, which play a crucial role in cosmic evolution during CD, leaving distinct imprints on a variety of observables.
Ionizing UV photons are likely the dominant driver of reionization (Sec.~\ref{SubSec:SFR}), whereas softer UV photons affect star formation and the spin state of hydrogen in the IGM: \Lya photons mediate spin-flip transitions in neutral hydrogen, modifying the 21-cm signal (Sec.~\ref{sub:T21}), and Lyman–Werner (LW) photons regulate star formation through radiative feedback (Sec.~\ref{SubSec:SFR}).
UV emission further serves as a key tracer of star formation via the UVLFs (Sec.~\ref{sub:UVLFs}).
In parallel, X-rays heat and partially ionize the IGM on large scales, thereby influencing both the 21-cm signal and reionization observables (Secs.~\ref{sub:T21} and ~\ref{Sec:IGMEvolve}). Moreover, the harder X-rays emitted by high-redshift sources free stream and contribute to the present day CXB (Sec.~\ref{sub:CXB}).

In what follows, we describe the default astrophysical model implemented in \Gcode\ and its connection to the evolution of the relevant radiation fields. The model specifies the formation of sources, their emission properties, and the propagation of the resulting radiation, from which we derive the production rate of ionizing UV photons as well as the global LW, \Lya, and X-ray intensities. While this default model provides a physically motivated and internally consistent description, the modular architecture of \Gcode \ (see App.~B) allows individual components to be modified independently without changing the overall framework; this flexibility is illustrated in Figs.~\ref{fig:T21} and \ref{fig:xHI}, where we implement the model of~\cite{Davies_2025_1}, which includes intrinsic variability in galaxy properties.

\subsection{Star Formation}\label{SubSec:SFR}
The first global quantity computed by \Gcode\ upon initialization is the star formation rate density (SFRD) of early stellar populations.  The default star formation model implemented in \Gcode \ closely follows~\cite{Munoz:2021psm,Park:2018ljd},
tracking two stellar populations corresponding to successive phases of star formation: Population III (Pop-III) stars, believed to be the first generation formed from pristine, metal-free gas at high redshifts; and Population II (Pop-II) stars, which are chemically enriched by the remnants of their Pop-III predecessors. The code can also be run in a single-population mode, with just Pop-II  star formation.

The global averaged SFRD of each stellar population \( i = \mathrm{II}, \mathrm{III} \), is given by
\begin{equation}\label{eq:SFRD}
\dot{\rho}^i_\star(z) = \int \frac{dn}{dM_{\rm h}}(M_{\rm h},z) \, f_{\rm gal}^{i}(M_{\rm h},z) \, \dot{M}^i_\star(M_{\rm h},z) \, dM_{\rm h} ,
\end{equation}
where we recall that \( dn/dM_{\rm h} (M_{\rm h},z)\) is the HMF, \( \dot{M}_\star^i(M_{\rm h},z) \) is the average star formation rate in galaxies hosted by halos of mass \( M_{\rm h} \), and \( f_{\rm gal}^i(M_{\rm h},z) \) is a smooth halo-mass filter that captures the cooling conditions required to form stars and, in the two-population scenario, also governs the transition between Pop-III- and Pop-II-dominated regimes.

By default, \texttt{\Gcode} adopts the Sheth–Tormen HMF~\cite{Sheth_1999}, computed with the public Python package \texttt{Colossus}~\cite{Diemer_2018} assuming a Planck18 cosmology~\cite{Planck:2018vyg}. Alternative HMF prescriptions implemented in \texttt{Colossus} can be selected through keyword arguments, and users may also override the \texttt{dndlnM} method within the SFRD module to supply a custom HMF.

\subsubsection*{Star formation rate within halos}

Following~\cite{Park:2018ljd, Munoz:2021psm,Davies_2025_1}, the average star formation rate within a halo is modeled as the stellar mass associated with the halo divided by a characteristic timescale
\begin{equation} \label{eq:SFR}
    \dot{M}^i_{\star}(M_{\rm h}, z) = \frac{M_\star^i(M_{\rm h})}{\eps_t \, H^{-1}(z)},
\end{equation}
where \( \eps_t \) is a dimensionless free parameter denoting the star formation timescale in units of the Hubble time. During matter domination, the halo dynamical time scales approximately with the Hubble time and is closely related to the gas depletion time in analytic models and numerical simulations (e.g.~\cite{Hopkins_2011,Hopkins_2013,Hopkins_2014,Dekel_2014}).

The stellar to halo mass relation is parametrized by
\begin{equation}
    M^i_\star(M_{\rm h}) = M_{\rm h} \, \frac{\Omega_b}{\Omega_m} \, f^i_\star(M_{\rm h}),
\end{equation}
where \( f_\star(M_{\rm h}) \) is the (integrated) star formation efficiency (SFE).
For Pop-II stars, the SFE as a function of halo mass is modeled as a double power law~
\begin{equation}
    f_\star^{\rm II}(M_{\rm h}) = \min \left[ \frac{F_\star^{\rm II}}{\left( \frac{M_{\rm h}}{M_{\rm p}} \right)^{\alpha^{\rm II}_\star} + \left( \frac{M_{\rm h}}{M_{\rm p}} \right)^{\beta^{\rm II}_\star}}, \ 1 \right] \,,
\end{equation}
where $\alpha^{\rm II}_\star<0$, $\beta^{\rm II}_\star>0$, $F_\star^{\rm II}$, and $M_{\rm p}$ are free parameters\footnote{To avoid strong parameter degeneracies, $f_\star^{\rm II}(M_{\rm h})$ is not normalized such that its peak value equals $F_\star^{\rm II}$. Consequently, depending on $\alpha_\star^{\rm II}$ and $\beta_\star^{\rm II}$, the normalization parameter $F_\star^{\rm II}$ can exceed unity while still satisfying the physical requirement $f_\star^{\rm II} \leq 1$.}. This form is motivated by the observed suppression of the luminosity and stellar mass function towards the high and low mass ends~\cite{Yang_2003,Moster_2010,Behroozi_2010,Harikane_2022,Behroozi_2019,Shuntov_2025}. The low-mass suppression is often interpreted as reflecting the effects of stellar
feedback in low-mass halos~\cite{dekel1986origin,Hopkins_2014}, while the physical origin of the high mass turnover is more debated and may be attributed to virial shocks~\cite{Birnboim_2003} and AGN feedback~\cite{Croton_2006,Benson_2003}, see also~\cite{Somerville_2015,Harikane_2022}.

Pop-III stars, on the other hand, are expected to form earlier in smaller halos, a hierarchy that is implemented through the halo mass filter $f_{\rm gas}^{\rm III}$. Therefore, their SFE is modeled as a single power law in halo mass
\begin{equation}
    f_\star^{\rm III}(M_{\rm h}) = \min \left[ F_\star^{\rm III} \left( \frac{M_{\rm h}}{10^7 M_\odot} \right)^{-\alpha^{\rm III}_\star}, \ 1 \right]\,.
\end{equation}

\subsubsection*{Cooling thresholds and the halo mass filter}
Star formation requires the baryonic gas to cool and fragment. In the absence of metals, cooling proceeds through radiative de-excitations of atomic or molecular hydrogen, provided the virial temperature of the halo is sufficiently high so that collisional excitations are efficient. Adopting the standard thresholds $T_{\rm vir} = 10^{3} {\rm K}$ and $10^{4} \ {\rm K}$ 
for molecular~\cite{Tegmark_1997, Bromm_2002} and atomic~\cite{barkana2001beginning} cooling, respectively, the corresponding minimum halo masses are~\cite{barkana2001beginning}
\begin{equation}
    M_{\rm mol}(z) = 5.3 \times 10^5  \left(\frac{1+z}{16}\right)^{-3/2}  M_\odot \,, \quad M_{\rm atom}(z) = 4.7 \times 10^7  \left(\frac{1+z}{16}\right)^{-3/2} M_\odot \,.
\end{equation}

Given the hierarchical nature of structure formation, the first stars likely formed in the smallest halos capable of cooling—those above the molecular cooling threshold. However, star formation in these halos is particularly fragile. LW photons emitted by stars can dissociate molecular hydrogen, suppressing 
cooling, and supersonic baryon–dark matter streaming velocities can inhibit gas collapse~\cite{Dalal_2010,Tseliakhovich_2011,machacek2001simulations,Fialkov_2014}. We incorporate these effects by defining the minimal halo mass for star formation as
\begin{equation}
M_{\rm cut}^{i}(z)=
\begin{cases}
M_0^i(z),
& M_0^i(z) \ge M_{\rm atom}(z), \\[6pt]

\max \left\{
    M_0^i(z),\;
    M_{\rm mol}(z)\,
    f_{\rm LW}(J_{\rm LW})\,
    f_{v_{\rm cb}}(v_{\rm bc})
\right\},
& M_0^i(z) < M_{\rm atom}(z).
\end{cases}
\end{equation}
where \( M_0 \) may be set either to the redshift dependent atomic cooling threshold, or to a user-specified fixed value. The factors 
\begin{equation}
    f_{\rm LW}(z) = 1+A_{\rm LW}\left(\frac{I_{LW}(z)}{10^{-21}{\rm erg \,s^{-1} \, cm^{-2} \, Hz^{-1} \, sr^{-1}}}\right)^{\beta_{\rm LW}} \quad {\rm and} \quad f_{v_{\rm bc}} = \left(1+A_{v_{\rm bc}} \frac{v_{\rm bc}}{v_{\rm rms}}\right)^{\beta_{v_{\rm bc}}} \,,
\end{equation}
quantify the impact of LW and streaming velocities respectively, with a function shape fit to simulations~\cite{machacek2001simulations, Kulkarni_2021,schauer2021influence,Munoz:2021psm}. Above, $I_{LW}$ is the mean intensity over the LW band, which is computed dynamically as described in Sec.~\ref{SubSec:Radiation}\footnote{The LW intensity is sourced by stars, thus resulting in a co-dependence between SFRD(z) and $I_{\rm LW}(z)$. The two quantities are evolved sequentially in redshift, starting at 
$z_\star=50$, above which star formation is assumed to be negligible.}, and $v_{\rm rms} = 29 \,{\rm km \,s^{-1}}$ is the root mean square relative baryon-CDM velocity. We compute $f_{v_{\rm bc}}$ assuming
$v_{\rm bc}=0.92 v_{\rm rms}$, corresponding to the mean of the Maxwell–Boltzmann distribution~\cite{Tseliakhovich_2010,Mu_oz_2019}.

Based on the above discussion, we define the following galaxy occupation functions~\cite{Munoz:2021psm}
\begin{equation}
    f_{\rm gal}^{\rm II}(M_{\rm h},z)
    = \exp\!\left(-\frac{M_{\rm cut}^{\rm II}}{M_{\rm h}}\right), \quad  f_{\rm gal}^{\rm III}(M_{\rm h},z)
    = \exp\!\left(-\frac{M_{\rm cut}^{\rm III}}{M_{\rm h}}\right)
      \exp\!\left(-\frac{M_{\rm h}}{M_{\rm cut}^{\rm II}}\right),
\end{equation}
which ensure a smooth suppression of star formation below the cooling threshold and a gradual transition from Pop-III to Pop-II dominance as increasingly massive halos form.

\subsubsection*{Photoheating feedback}

In addition to the feedback mechanisms described above, UV ionizing radiation produced by stars during the EoR photoheats and evaporates the gas in small structures, suppressing subsequent star formation in halos with \( M_{\rm h} \lesssim 10^9 \, M_\odot \)~\cite{Sobacchi_2013,Sobacchi_2013b}. This effect extends above the atomic cooling threshold and 
could potentially increase both \( M_{\rm cut}^{II} \) and \( M_{\rm cut}^{III} \). However, as shown in~\cite{katz2025closingpopiiistarsconstraints}, for SFE models consistent with UVLF observations, the Pop II SFRD is already dominated by halos with masses $M_{\rm h} > 10^9 \ M_{\odot}$. We therefore neglect the effect of photoheating feedback on Pop-II stars. In contrast, the SFRD of Pop-III stars, which only form in $M_{\rm h}<M_{\rm cut}^{\rm II}$ halos,
could be strongly suppressed. 

A full treatment of photo-heating feedback on Pop-III star-forming halos would require tracking the evolution of ionization fronts. \Gcode\ instead approximates this effect by rescaling the Pop III SFRD with the volume fraction that has not yet been ionized by UV photons:
\begin{equation}\label{eq:photoheat_feedback}
    \dot{\rho}_\star^{\rm III}(z)\rightarrow \dot{\rho}_\star^{\rm III}(z)(1-Q_{\rm HII}(z)) \,,
\end{equation}
where the evolution of the UV ionized fraction, $Q_{\rm HII}$, is described in Sec.~\ref{SubSec:xHI}. 
Photo-heating feedback can be toggled on or off in \Gcode. 

Overall, when all feedback effects are included, this star formation model takes twelve parameters: \( \eps_t \), \( F_\star^i \), \( \alpha^i_\star \), \( \beta^{\rm II}_\star \), \( M_0^i \), \( A_{\rm LW} \), \( \beta_{\rm LW} \), \( A_{v_{\rm cb}} \), and \( \beta_{v_{\rm cb}} \). We note that while the parameters \( F_\star^i \), and \( \eps_t\) are mostly degenerate, the physical constraint $f_\star(M_{\rm h})\leq 1$ breaks this degeneracy in extreme models.

\subsection{Radiation: Emission and Propagation}\label{SubSec:Radiation}

Observations at lower redshift and population synthesis models suggest that the dominant source of X-ray emission in star-forming galaxies at $z \gtrsim 5$ are high-mass X-ray binaries (HMXBs)~\cite{Grimm:2002ta,ranalli20032,Gilfanov:2003bd,Fabbiano:2005pj,Mineo:2011id,Fragos:2013bfa}, while massive stars are likely the primary UV emitters~\cite{Madau:2014bja,Gessey_Jones_2022,leitherer1999starburst99,Stanway_2015}.
In both cases, sources are expected to be short-lived, implying that the total emission is directly proportional to the star formation rate. Consequently, the global (intrinsic) specific photon emissivity — defined as the number of photons emitted per unit time, frequency and comoving volume — of Pop-i sources can be expressed as
(for a full derivation and validation, including its applicability to Pop-III stars, see App.B of~\cite{katz2025closingpopiiistarsconstraints})
\begin{equation} 
    \epsilon^{i}(z,\nu) = \frac{\dot{\rho}_\star^{i}(z)}{\mu_{\rm b}} \left\langle \frac{dN}{d\nu} \right\rangle^{i}
    \label{eq:emissivity}
\end{equation}
where $\mu_{\rm b}$ is the mean baryon mass, and $\langle dN/d\nu \rangle^{i}$ is the average emission spectrum per baryon in Pop-i sources.

Once emitted, photons propagate through the dense ISM of the host galaxy and the diffuse IGM, interacting with the surrounding gas. Depending on their energy, these photons may excite, ionize, or heat the medium, thereby altering both the physical state of the gas, which we discuss in Sec.~\ref{Sec:IGMEvolve}, and the emergent radiation spectrum, which is the focus here. It is convenient to characterize the radiation fields in terms of the global specific photon intensity, defined as the number of photons per unit time, frequency, area, and solid angle (see~\cite{Haardt_1996} for derivation)
\begin{equation}
    J(z,\nu) = \frac{c}{4\pi} (1+z)^2\int_z^\infty \frac{\sum_{i=II,III} \epsilon^i\!\left( z',\nu \frac{1+z'}{1+z}\right)}{H(z')} \, \mathcal{T}(\nu,z,z') \, dz',
    \label{eq:flux}
\end{equation}
where $H(z)$ is the Hubble parameter, and $\mathcal{T}(\nu,z,z')$ is a generalized transfer function representing the probability that a photon emitted at redshift $z'$ with frequency $\nu' = \nu(1+z')/(1+z)$ contributes to the radiation field at redshift $z$ and frequency $\nu$. For a purely absorbing medium, $\mathcal{T}(\nu,z,z') = e^{-\tau(\nu,z,z')}$, where $\tau(\nu,z,z')$ is the optical depth. The photon emissivity and intensity should not be confused with the corresponding energy quantities related through 
$ \mathcal{E}^i(z,\nu) =   \eps^i(z,\nu)h\nu$ and $I(z,\nu) =  J(z,\nu)h\nu$ respectively\footnote{We caution that simply replacing the photon emissivity in Eq.~\eqref{eq:flux} with the energy emissivity $\mathcal{E}(z',\nu')$ does not yield the correct energy intensity $I(z,\nu)$. An additional $\left( \frac{1+z}{1+z'} \right)$ factor is required to account for the redshifting of photon energies from $\nu'$ to $\nu$.}.

In the remainder of this section, we describe the default models for the spectra and propagation of photons across the UV and X-ray regimes. Although the intrinsic UV and X-ray spectra of a stellar population are, in principle, linked through the initial mass function (IMF) and the assumed spectral models of individual sources, enforcing this correlation introduces significant modeling complexity and uncertainty, primarily due to the poorly constrained IMF at high redshift (see~\cite{Gessey_Jones_2025}, where they instead demonstrate how an assumed UV–X-ray correlation can be used to determine the stellar IMF at CD). We therefore treat the UV and X-ray spectra independently. Within the UV band, we maintain a consistent emission spectrum but distinguish between ionizing and non-ionizing (Lyman-band) photons, given their qualitatively different interactions with the surrounding gas.

\subsubsection{Non-Ionizing UV Radiation: \Lya and LW Intensities}
\label{sec:Nonion_UV}

The Lyman-band spectrum in \Gcode \ is modeled as a broken power law with breaks at each Lyman transition, and is by default fitted to the population synthesis models~\cite{leitherer1999starburst99} for Pop-II and~\cite{bromm2001generic} for Pop-III stars, following~\cite{barkana2001beginning}. To ensure consistency across the entire UV spectrum, the normalization over the Lyman band is scaled with the total number of ionizing photons emitted per baryon, \(N_{\gamma/{\rm b}}^{i}\), which is a user specified parameter. Throughout this work, we will adopt the fiducial values \(N_{\gamma/{\rm b}}^{\rm II} = 5000\) and \(N_{\gamma/{\rm b}}^{\rm III} = 44021\), assumed in 21cmFAST~\cite{Munoz:2021psm}. For flexibility, the spectral shape is defined via an external file, which the user can easily modify as needed. However, we note that, due to the narrowness of the Lyman band, the exact spectral shape has little impact on derived observables, with the results being mostly sensitive to the normalization of the spectrum. The Lyman-band spectra, combined with the SFRD of Eq.~\eqref{eq:SFRD} defines the pop-i Lyman-band emissivities, $\eps^i_{\rm Ly}(z,\nu)$ through Eq.~\eqref{eq:emissivity}, from which the corresponding \Lya and LW intensities are computed as described below.

Once emitted, Lyman-band photons free-stream until they redshift to the energy of the nearest Lyman line, where they are immediately absorbed due to the high hydrogen density~\cite{Barkana:2004vb,Hirata:2005mz}. The excited hydrogen atoms then cascades to one of the fine $n=2$ states, either $2s$ ($l=0$) or $2p$ ($l=1$), before eventually relaxing to the ground state. A transition from the $2s$ state to the ground state is only allowed via a three-photon decay, whereas de-excitation from the $2p$ state produces a \Lya photon.
With this picture in mind, the \Lya flux at redshift $z$, that is sourced by stellar emission, can be computed by summing the contributions from all emitted photons that reach a $n \geq 2$ Lyman line at the redshift of interest\footnote{In practice, following~\cite{Barkana:2004vb}, we cut the sum at $n=23$. Increasing this value has a negligible effect on the results, as this choice already encompasses $99.8\%$ of the Lyman band.
}, 
\begin{equation}
    \label{eq:J_alpha}
    J^\star_\alpha(z) = \sum_{n=2}^{\infty} P_n J_n(z, \nu_n) \,,
\end{equation}
where $\nu_n$ is the frequency of the Lyman-$n$ resonance, and the factors $P_{n}$ correspond to the probability that a cascade starting at Lyman level $n$ terminates with the production of a \Lya photon; see~\cite{Hirata:2005mz} for tabulated values. The fluxes, $J_{n}(z,\nu)$ are computed according to Eq.~\eqref{eq:flux} with the Lyman band emissivity describe above, and a corresponding heavyside step transfer function
\begin{equation}\label{eq:Transfer_Lya}
    \mathcal{T}_n(\nu,z,z') = \Theta \left(\frac{1+z}{1+z'}\nu_{n+1} - \nu  \right) \,.
\end{equation}
This transfer function accounts for the opaqueness of Lyman lines, ensuring that photons are not double counted.

In addition to stellar non-ionizing UV radiation, which is expected to be the dominant source of \Lya\ photons during Cosmic Dawn, X-rays provide an additional contribution, $J_\alpha^{\rm X}(z)$, through secondary excitations. This contribution is included in \Gcode{} as described in Sec.~\ref{subsubsec:Xrays}.
Finally, absorption and re-emission of photons near the Doppler-broadened \Lya line further modify the spectral profile around line center~\cite{Hirata:2005mz}; this radiative transfer effect is accounted for separately, as discussed in Sec.~\ref{sub:T21}.

The energy intensity at the LW band is computed in a similar, but opposite manner to $J^\star_\alpha(z)$. Photons that redshift into a Lyman resonance cascade to energies below the LW band ($11.2$--$13.6~{\rm eV}$). Hence, the band-averaged LW energy flux at redshift $z$ is obtained by summing over all photons with energies $E \geq 11.2~{\rm eV}$ that have yet to reach a Lyman resonance
\begin{equation}
    I_{\rm LW}(z) = 
    \frac{\sum_{n=2}^\infty \int_{\max(\nu^{\rm min}_{\rm LW},\nu_n)}^{\nu_{n+1}}  I_n(z,\nu)\, d\nu}{\Delta \nu_{\rm LW}} 
    \approx 
    \frac{c}{4\pi}(1+z)^3 \int_z^{\infty} dz' \, \frac{\bar{\mathcal{E}}_{\rm LW}(z')}{(1+z')H(z')} \,
    f_{\rm mod}\!\left(\frac{1+z'}{1+z}\right)\,.
    \label{eq:JLW}
\end{equation} 
Above, $h\nu^{\rm min}_{\rm LW}=11.2~{\rm eV}$ and $h\Delta\nu_{\rm LW}=2.4~{\rm eV}$ denote the lower bound and width of the LW band, while $I_n(z,\nu)$ are the energy intensities corresponding to the number intensities $J_n(z,\nu)$ introduced in Eq.~\eqref{eq:J_alpha}. 
To avoid the computationally expensive double integral implicit in the first equality of Eq.~\eqref{eq:JLW}, we exploit the narrowness of the LW band and approximate the energy emission spectrum as flat, replacing the energy emissivity by its average over the LW band, $\bar{\mathcal{E}}_{\rm LW}$.
The approximation is applied after the second equality, and allows to evaluate the energy integral independently of 
astrophysical parameters~\cite{ahn2009inhomogeneous} (see also~\cite{Fialkov_2013,qin2020tale}), defining
\[
f_{\rm mod}\!\left(\frac{1+z'}{1+z}\right) 
= \frac{1}{\Delta \nu_{\rm LW}}
\sum_{n=2}^{\infty} 
\int_{\max(\nu^{\rm min}_{\rm LW},\nu_n)}^{\nu_{n+1}} 
\mathcal{T}_n(\nu,z,z') \, d\nu \,.
\]

The Ly$\alpha$ and LW fluxes derived here play a significant role in determining the 21-cm signal (Sec.~\ref{sub:T21}) and the SFRD (Sec.~\ref{SubSec:SFR}).

\subsubsection{Ionizing UV Radiation}

Unlike the other radiation fields, \Gcode{} does not explicitly model the ionizing UV spectrum. Instead, the ionizing photon production of Pop-\(i\) stars is parameterized by the average number of ionizing photons emitted per baryon in stars, \(N_{\gamma/{\rm b}}^{\,i}\), which also determines the normalization of the Lyman-band spectrum.

Of the ionizing UV photons produced, a fraction is absorbed by dense gas in their host galaxies, where rapid recombinations maintain large neutral fractions, preventing these photons from contributing to cosmic reionization. Galaxy formation simulations indicate that the resulting escape fraction may depend on the mass of the host halo~\cite{Paardekooper_2015, Ma_2020}. Following~\cite{Park:2018ljd}, we model the escape fraction as
\begin{equation}
    \label{eq:Fesc}
    f_{\rm esc}^{\,i}(M_{\rm h})
    \;=\; \min \left[
    F_{\rm esc}^{\,i}\!\left(\frac{M_{\rm h}}{M_{i}}\right)^{-\alpha_{\rm esc}^{\,i}},\ 1\right],
\end{equation}
where \(M_{i}=10^{10}\,M_\odot\) for Pop-II and \(M_{i}=10^{7}\,M_\odot\) for Pop-III galaxies. The distinct parameter sets for Pop-II and Pop-III systems account for possible differences in their structure and chemical composition. While  Eq.~\eqref{eq:Fesc} provides a relatively flexible dependence on halo mass, we note that some simulations suggest more complex behaviors (e.g.~\cite{Ma_2020}). Accounting for the escape fraction, the corresponding rate at which ionizing photons are emitted to the IGM is 
\begin{equation}
    \label{eq:Nion}
    \dot{N}_{\rm ion}^{\rm UV}
    \;=\; \sum_{i\in\{\mathrm{II,III}\}}N_{\gamma/{\rm b}}^{\,i}
    \frac{1}{\rho_{\rm b}^{0}}
    \int \frac{dn}{dM_{\rm h}}(M_{\rm h},z) \, f_{\rm gal}^{i}(M_{\rm h},z) \, \dot{M}^i_\star(M_{\rm h},z) \,
    \,
    f_{\rm esc}^{\,i}(M_{\rm h})\,
    dM_{\rm h}\,,
\end{equation}
where we have implicitly assumed ionizing sources are short lived, such that ionizing photon production traces the instantaneous star-formation rate~\cite{leitherer1999starburst99,Stanway_2015}.

Given the modest excess energies of stellar UV photons, we approximate each ionizing UV photon as producing a single ionization event. This approximation does not apply to X-ray–induced ionization, which involves energetic photoelectrons and is discussed in Sec.~\ref{subsubsec:Xrays}. 
Moreover, because the photoionization cross section is enhanced at low energies, the time between the emission of an ionizing UV photon and the subsequent photoionization event is typically short compared to variations in the SFRD, which occur on timescales of order 
$\gtrsim100 \ {\rm Myr}$ during reionization (see Eq.~\eqref{eq:SFR} and App.~B of~\cite{katz2025closingpopiiistarsconstraints})\footnote{This approximation may break toward the end of reionization, when ionized regions grow to hundreds of comoving megaparsecs (see, e.g., \cite{Lin_2016,Lu_2025}) and UV photons can propagate over significant distances before interacting with neutral hydrogen}. Consequently, $\dot{N}^{\rm UV}_{\rm ion}(z)$ can be identified as the UV induced ionization rate at redshift $z$.

Overall, for a given star-formation history, the ionizing UV emission and its propagation are described by three UV-ionization parameters for each stellar population: $N_{\gamma/b}^{\,i}$, determined by the ionizing sources, and $(F_{\rm esc}^{\,i},\,\alpha_{\rm esc}^{\,i})$, which encapsulate the properties of their host galaxies.

\subsubsection{X-rays}
\label{subsubsec:Xrays}

In agreement with population-synthesis models of low-metallicity galaxies~\cite{Fragos:2013bfa} and observations of local galaxies whose X-ray emission is dominated by ultraluminous non-AGN sources~\cite{Wik_2014,Yukita_2016,Garofali_2020,lehmer20150330kevspectrapowerful}—interpreted as X-ray binaries—\Gcode{} models the $E\leq 30 \  {\rm keV}$ X-ray spectra as a double power law
\begin{equation}\label{eq:XraySpectrum}
    \left\langle \frac{d N_X}{d\nu} \right\rangle =
    \begin{cases}
        \frac{1}{\nu_0} \left( \frac{\nu}{\nu_0} \right)^{-1 - \alpha^{\rm s}_X}, & h\nu \leq E_{\rm break}, \\[6pt]
        \frac{1}{\nu_0} \left( \frac{E_{\rm break}}{h\nu_0} \right)^{\alpha^{\rm h}_X - \alpha^{\rm s}_X} \left( \frac{\nu}{\nu_0} \right)^{-1 - \alpha^{\rm h}_X}, & h\nu > E_{\rm break}\,,
    \end{cases}
\end{equation}
where \(\alpha^{\rm s}_X\), \(\alpha^{\rm h}_X\) and \(E_{\rm break}\) are user specified parameters. 
To remain consistent with standard 21-cm modeling, and to define a physically intuitive normalization parameter, \(\nu_0\) is fixed to match the $<2 \ {\rm keV}$ integrated X-ray luminosity to SFR ratio, that is 
\begin{equation}\label{eq:XrayNorm}
\frac{L_{\left[E_\text{min},2\,\mathrm{keV}\right]}}{\dot{M}_\star}
    = \frac{1}{\mu_b} \int_{E_{\rm min}}^{2\,\mathrm{keV}} E \left\langle \frac{dN_X}{dE}\right\rangle dE \, ,
\end{equation}
where $\frac{L_{\left[E_\text{min},2\,\mathrm{keV}\right]}}{\dot{M}_\star}$ and $E_{\rm min}$ are user defined. The right hand side of Eq.~\eqref{eq:XrayNorm} follows from $\frac{dL}{dE}=\frac{\dot{M}_\star}{\mu_b} E  \left\langle \frac{dN_X}{dE}\right\rangle$, consistent with the short lifetime approximation for X-ray sources (see discussion above Eq.~\eqref{eq:emissivity}).

Both observations and theoretical models indicate that the integrated X-ray luminosity-to-SFR ratio of HMXBs increases toward lower metallicities~\cite{Fragos:2013bfa,brorby2016enhanced,lehmer2021metallicity}. In the present version of \Gcode{}, this dependence is only partially captured by allowing different luminosity-to-SFR ratios for Pop~II and Pop~III galaxies, while a continuous metallicity-dependent prescription will be implemented in future work~\cite{Xray}.

Using the X-ray spectrum defined above, \Gcode \ computes the X-ray photon intensity, $J_{\rm X}$,
following Eqs.~\eqref{eq:emissivity} and \eqref{eq:flux}, accounting for line-of-sight absorption through
\begin{equation}
    \mathcal{T}(\nu,z,z') = \Theta(h\nu-E_{\rm min}) \, e^{-\tau(\nu,z,z')} \,.
\end{equation}
Here, the Heaviside step function and the exponential respectively describe absorption within the host galaxy’s ISM and the IGM.
While photoionization dominates the attenuation in both regimes, uncertainties in the average ISM density and composition necessitate additional modeling, which is encapsulated in the parameter $E_{\rm min}$\footnote{$E_{\min}$ can be thought of as a measure for the average ISM column density, corresponding to the energy at which the ISM optical depth reaches unity~\cite{Park:2018ljd}.}. In contrast, \Gcode \ self consistently evolves the [HI,HeI,HeII] abundances in the IGM, allowing for a direct computation of the optical depth within it
\begin{equation}
    \tau(\nu,z,z') = \sum_{i = {\rm HI}, {\rm HeI}, {\rm HeII}} 
    \int_{z}^{z'} \frac{dz''}{H(z'')(1+z'')} \, n_i(z'') \, \sigma_i\!\left( \nu'' \right) \,,
\end{equation}
where \(\sigma_i(\nu'')\) is the photoionization cross section at 
\(\nu'' \equiv \frac{1+z''}{1+z} \nu\)~\cite{Verner:1996th}. Overall, $J_{\rm X}$ is defined by six X-ray parameters: $E_{\rm min}$, $\alpha_{\rm h}$, $\alpha_{\rm s}$, $E_{\rm break}$, and the soft band luminosity to SFR ratio for both populations.

Lastly, we compute the \Lya\ flux generated by X-ray interactions with the IGM. X-ray induced photoionizations in the IGM generate energetic photoelectrons whose subsequent interactions heat, ionize, and excite the IGM.  Hydrogen atoms excited by these secondary interactions decay through radiative cascades (see Sec.~\ref{sec:Nonion_UV}), potentially producing \Lya photons. 

With the specific X-ray photon intensity $J_{\rm X}(z,\nu)$ derived above, the volumetric rate of energy deposition into the IGM by photoionizations of species $i\in{\mathrm{HI},\mathrm{HeI},\mathrm{HeII}}$—excluding the ionization threshold energy—is given by
\begin{equation}
\label{eq:E_dense}
    u_i(z) = 4\pi n_i(z)  \int_{\nu_{\rm ion}^i}^{\infty} (h\nu - h\nu_{\rm ion}^i)\, \sigma_i(\nu)\, J_X(z,\nu) \, \, d\nu \,,
\end{equation}
where $h\nu_{\rm ion}^i$ denotes the ionization threshold of specie $i$. 
Assuming a quasi–steady balance between X-ray injection and Hubble-flow escape from line center, the X-ray contribution to the specific number flux at the \Lya frequency are computed as~\cite{Pritchard:2006sq} 
\begin{equation}
    J_{\alpha}^{\rm X}(z) = \frac{c}{4\pi}\frac{\dot{n}_\alpha^{\rm X}(z)}{H(z)\nu_\alpha}\,,
\end{equation}
with
\begin{equation}
    \dot{n}_\alpha^{\rm X}(z) = \frac{1}{h\nu_\alpha}\sum_iu_i(z)\eps_{ \alpha}(x_{\rm HI}^{\rm out})
\end{equation} 
where $\eps_{ \alpha}(x_{\rm HI})$ is the fraction of the initial photoelectron energy that goes to producing \Lya photons. $\eps_{ \alpha}(x_{\rm HI})$, along with the corresponding fractions for heating and HI ionization, denoted by $\eps_{\rm heat}(x_{\rm HI})$ and $\eps_{\rm ion}^{\rm HI}(x_{\rm HI})$ respectively, are tabulated in~\cite{Furlanetto_2010}. Notice that the energy deposition fractions should be evaluated according to the average neutral hydrogen fraction outside of UV ionized bubbles, which we write as $x_{\rm HI}^{\rm out}$ (see Sec.~\ref{SubSec:xHI})

In principle, the deposition fractions depend on the primary photon energy. However, this dependence saturates for $E \gtrsim 100 \ {\rm eV}$~\cite{Furlanetto_2010,efficiency}, well below the adopted X-ray escape energy $E_{\rm min}$ inferred from simulations~\cite{Das:2017fys}. Although emitted photons can redshift below $E_{\rm min}$, the high IGM opacity to soft ionizing photons~\cite{McQuinn_2012} prevents any significant contribution at $E < 100 \ {\rm eV}$. We therefore use the high-energy asymptotic deposition fractions of~\cite{Furlanetto_2010}

\section{IGM Evolution}\label{Sec:IGMEvolve}

Throughout this section, we review the evolution of the IGM in \Gcode{} for the standard $\Lambda$CDM settings. In this framework, \Gcode{} self-consistently solves the coupled evolution of the neutral hydrogen fraction, $x_{\rm HI}$, and the gas kinetic temperature, $T_{\rm k}$, over a wide redshift range. By default, the evolution is initialized at $z=1200$ with $x_{\rm HI}=1$ and $T_{\rm k}=T_\gamma$ and terminated at $z=6$, thereby capturing hydrogen recombination, CD, and the EOR. More generally, \Gcode{} allows the user to initialize the evolution at any redshift after the end of helium recombination and before the onset of CD, $1800 \leq z \leq 35$~\cite{katz2025closingpopiiistarsconstraints,Ali-Haimoud:2010hou}.

\subsection{Ionization History}\label{SubSec:xHI}
\Gcode{} tracks the ionization state of the IGM across cosmic time. Prior to CD, helium is assumed to be fully recombined, while during CD and the EOR we adopt the standard approximation~\cite{efficiency,Mesinger:2010ne} that the singly ionized helium fraction, $x_{\rm HeII} \equiv n_{\rm HeII}/n_{\rm He}$, equals that of hydrogen, $x_{\rm HII}$, neglecting doubly ionized helium entirely. In what follows we describe the evolution of $x_{\rm HII}$.

During the first phase of the evolution, prior to the onset of star formation at \(z=z_\star\), the CMB is the only significant radiation background. Even at the highest redshifts of interest, the CMB spectrum contains a negligible number of photons capable of ionizing hydrogen directly from its ground state. The evolution of the ionized fraction is therefore governed by the dynamics of the hydrogen atomic level populations, which we follow using the three-level atom approximation (see, e.g.,~\cite{Ali-Haimoud:2010hou}):
\begin{equation} \label{eq: xe evolution}
    \frac{dx_{\rm HII}}{dz} (z>z_\star)=-\frac{1}{H(z)(1+z)} C_p 
    \left[ 4\left(1-x_{\rm HII}\right)\beta_B(T_\gamma) e^{-\frac{h\nu_{\alpha}}{ k_{\rm B}T_\gamma} }  - \alpha_{\rm B}(T_{\rm k})n_{\rm e}(z) x_{\rm HII}\right]\, ,
\end{equation}
where $C_p(H,T_{\rm k},n_{\rm HI})$ is the Peebles factor~\cite{peebles1968recombination}, representing the probability that an excited atom in the $n=2$ state decays to the ground state before being photoionized. The coefficients $\alpha_{\rm B}(T_{\rm k})$ and $\beta_{\rm B}(T_{\gamma})$ denote the case-B recombination and photoionization rates respectively, as given in~\cite{Ali-Haimoud:2010hou}. 
Recombination is commonly described in two limiting cases. In case B, recombinations directly to the ground state are ineffective because the emitted ionizing photon is rapidly reabsorbed by a neighboring hydrogen atom, whereas in case A recombinations to all atomic levels contribute effectively. The first is more suited for the dense and neutral IGM at these redshifts.

By CD, the CMB spectrum has redshifted such that CMB-induced ionizations are negligible, while ionization is subsequently driven by the emergence of astrophysical sources. To ensure a smooth transition, \Gcode{} begins incorporating astrophysically driven ionization already at 
$z_\star=50$, where its contribution remains small even for extreme astrophysical models~\cite{katz2025closingpopiiistarsconstraints}. 
As star formation accelerates, photoionizations from stellar UV and X-ray emission rapidly outpace recombinations. 
Because the photoionization cross section declines steeply with photon energy, radiation in these two bands imprints distinct spatial morphologies: UV photons have extremely short mean free paths, producing localized ionization around astrophysical sources that expands outward, whereas X-rays propagate over much longer distances, generating a more homogeneous ionization background.
The combined effect is modeled as
\begin{equation}
x_{\rm HII}(z\leq z_\star) = \left(1-Q_{\rm HII}(z)\right)x^{\rm out}_{\rm HII}(z) + Q_{\rm HII}(z) \,
\end{equation}
where $Q_{\rm HII}(z)$ denotes the volume fraction of the universe filled by UV-driven ionized bubbles at redshift $z$, and $x^{\rm out}_{\rm HII}(z)$ represents the ionized fraction in regions of the IGM not yet encompassed by such bubbles.

To capture the large-scale impact of UV ionization without resolving individual ionization fronts, \Gcode{} adopts a global, volume-averaged approach. Despite the simplifying assumptions inherent in such a treatment, we find excellent agreement with the semi-analytical simulation \textsc{21cmFAST}, as demonstrated in Fig.~\ref{fig:xHI}. In this framework, the evolution of the ionized volume filling fraction is governed by~\cite{madau1999radiative,Mason_2019}
\begin{equation}\label{eq:FillingFraction}
    \frac{dQ_{\rm HII}}{dz}
    \;=\; -\frac{1}{H(z)(1+z)} \left[
    \dot{N}_{\rm ion}^{\rm UV}(z) - 
    C\alpha_{\rm B}\!\left(T=10^{4}\,{\rm K}\right)n_{\rm e}(z)Q_{\rm HII}\right]\,,
\end{equation}
where the two competing terms on the right-hand side describe, respectively, the UV induced ionization rate, given in Eq.~\eqref{eq:Nion}, and the global recombination rate.
Here, $C \equiv \langle n_{\rm e}^{2} \rangle / \langle n_{\rm e} \rangle^{2}$ is the clumping factor (not to be confused with the Peeble's factor in Eq.\eqref{eq: xe evolution}), which we fix to $C = 3$ based on numerical simulations~\cite{shull2011criticalstarformationratesreionization,Finlator_2012,Kaurov_2015}. The case-B recombination coefficient is evaluated at the atomic cooling threshold temperature (see Sec.~\ref{SubSec:SFR}), which is the typical temperature expected for photoheated gas.

Outside of UV-ionized regions, ionization is dominated by X-rays, such that
\begin{equation}
    \frac{dx_{\rm HII}^{\rm out}}{dz}=-\frac{1}{H(z)(1+z)} 
    \left[ \Gamma_{\rm ion}^{\rm X}(z,x_{\rm HII}^{\rm out})  - C\alpha_{\rm A}(T_{\rm k})n_{\rm e}(z) x_{\rm HII}^{\rm out}\right]\, ,
\end{equation}
where we now use the case-A recombination coefficient, $\alpha_{\rm A}(T_{\rm k})$~\cite{Abel_1997}, which is more appropriate for the dilute IGM during CD\footnote{The precise redshift of the transition from case-B to case-A recombination in the IGM has a negligible impact on the global ionization history.}. Because X-ray photons are far more energetic than the hydrogen ionization threshold, X-ray–induced photoionizations generate energetic secondary electrons that undergo further scattering, producing multiple ionizations. The total X-ray ionization rate is therefore given by
\begin{equation}
    \Gamma^{\rm X}_{\rm ion}(z,x_{\rm HI}^{\rm out}) = \frac{1}{n_{\rm H}(z)} \left[ 4\pi n_{\rm HI}(z) \int_{\nu_{\rm min}}^{\infty}d\nu \sigma_{\rm HI}(\nu)J_{\rm X}(z,\nu)  \ + \ \sum_i \frac{u_i(z)}{h\nu_{\rm ion}^{\rm HI}}\epsilon_{\rm ion}^{\rm HI}(x_{\rm HI}^{\rm out}) \right] \,,
    \label{eq:GammaXIon}
\end{equation} 
where the first term on the RHS counts the primary HI photoionization events and the second term counts contributions from secondary events. 
We remind the reader that $J_{\rm X}(z,\nu)$ denotes the specific global X-ray photon intensity, $u_i(z)$ is the total energy deposition rate per unit volume from secondary processes triggered by primary photoionizations of IGM species $i \in \{\mathrm{HI}, \mathrm{HeI}, \mathrm{HeII}\}$, $\epsilon_{\rm ion}^{\rm HI}(x_{\rm HI}^{\rm out})$ is the fraction of this energy deposited into $\mathrm{HI}$ ionizations, and $E_{\rm min} = h\nu_{\rm min}$ is the minimal energy at which X-ray photons escape the ISM of their host galaxy and reach the IGM; all of these quantities are modeled in Sec.~\ref{subsubsec:Xrays}.

\subsection{Kinetic Temperature Dynamics}\label{SubSec:Tk}

The evolution of the kinetic temperature of the IGM in \Gcode{} is solved simultaneously with the ionization rate, and when ran on $\Lambda$CDM settings is governed by
\begin{equation} \label{eq:Tk evolution}
    \frac{dT_k}{dz} = \frac{2T_k}{1+z} - \frac{T_{\rm k}}{1+x_{\rm e}+f_{\rm He}}\frac{dx_{\rm e}}{dz} - \frac{2}{3k_{\rm B}}\frac{1}{H(z)(1+z)}\left(\dot{Q}_{\rm Comp} + \dot{Q}_X + \dot{Q}_{\rm Ly\alpha} + \dot{Q}_{\rm CMB}\right),
\end{equation}
where $f_{\rm He}\equiv n_{\rm He}/n_{\rm H}$. The first term describes adiabatic cooling due to cosmic expansion, the second accounts for the change in the number of particles sharing the thermal energy as recombination and ionization proceed, and the \(\dot{Q}\) terms denote heating rates per baryon arising from the processes described below.

Prior to Cosmic Dawn, the only relevant additional heat transfering process is Compton scattering between CMB photons and the ionized component of the IGM. The corresponding heat transfer rate is given by (see, e.g.,~\cite{Liu_2020})
\begin{equation}
    \dot{Q}_{\rm Comp} = \frac{3}{2} \Gamma_{\rm Comp}(T_\gamma - T_k), \quad \text{where}, \quad \Gamma_{\rm Comp} = \frac{x_e}{1 + x_e + f_{\rm He}} \frac{8 \sigma_T u_\gamma}{3m_ec}
\end{equation}
is the Compton scattering rate per baryon, $u_\gamma$ is the energy density of the CMB and $m_{\rm e}$ is the electron mass. 
The remaining heat transferring processes are initially neglected and are introduced at \(z_\star = 50\), where their contribution is still small, ensuring a smooth transition into the Cosmic Dawn.

Once star formation initiates, soft ($E \lesssim 2\,\mathrm{keV}$) X-ray photons  may provide a significant heating channel (see Sec.~\ref{subsubsec:Xrays}). These photons interact with the IGM predominantly through photoionization, with the absorbed energy distributed via secondary processes, yielding a heating rate of
\begin{equation}
    \dot{Q}_{\rm X} = \frac{\eps_{\rm heat}(x_{\rm HI}^{\rm out})}{n_{\rm b}(z)}\sum_{i={\rm HI, HeI, HeII}} u_i(z) \,.
    \label{eq:Xheat}
\end{equation}

Typically subdominant to X-ray heating, the two remaining terms in Eq.~\eqref{eq:Tk evolution} describe heat exchange between hydrogen atoms and radiation through scattering processes involving \Lya and 21-cm CMB photons. These interactions exchange energy with the gas through two channels: photon recoil, which directly transfers kinetic energy to atoms, and excitations/ de-excitations of the hyperfine levels, which exchange energy with the internal spin reservoir. While \cite{Venumadhav:2018uwn} argued that \Lya-mediated coupling between the spin and kinetic temperatures can, in principle, allow energy associated with hyperfine transitions driven by the CMB to contribute significantly to gas heating, \cite{Meiksin:2021cuh} pointed that if the hyperfine levels are in steady state, the spin contributions from \Lya and CMB interactions cancel. Indeed, hyperfine transitions occur on timescales much shorter than the evolution of the radiation fields, so hyperfine populations are therefore in steady state, and the only net contribution to the kinetic temperature evolution arises from photon recoil and is given by~\cite{Meiksin:2021cuh} 
\begin{equation}
    \dot{Q}_{\rm Ly\alpha} = \frac{x_{\rm HI}}{1+x_{\rm e}+x_{\rm He}}\Gamma_\alpha  \frac{(h\nu_\alpha)^2}{m_{\rm H}c^2}\left( 1-\frac{T_{\rm k}}{T_\alpha} \right) \quad , \quad \dot{Q}_{\rm CMB} = \frac{x_{\rm HI}}{1+x_{\rm e}+x_{\rm He}}\Gamma_{\rm \gamma} \frac{E_{21}^2}{m_{\rm H}c^2}\left( 1-\frac{T_{\rm k}}{T^{\rm L}_{\rm CMB}} \right) \,.
\end{equation}

Similarly to Eq.~\eqref{eq:LyaRate}, we compute the total \Lya scattering rate per hydrogen atom as 
$\Gamma_\alpha = 6\pi\lambda_\alpha^2\gamma_\alpha J_\alpha\int \frac{J(\nu)}{J_\alpha} \bar{\phi}(\nu)d\nu$, where the frequency integral is analogous to $\tilde{S}_\alpha$, but uses the spin averaged \Lya line profile~\cite{Hirata:2005mz}, $\bar{\phi}_\alpha(\nu) = \frac{1}{4}\left(\phi_{00}+\phi_{01}\right) + \frac{3}{4}(\phi_{10}+\phi_{11})$. 
For the CMB we assume a flat spectrum with delta resonance, which in the limit
$T_{\rm s}\gg E_{21}$ gives a  scattering rate per hydrogen atom $\Gamma_\gamma = \frac{3}{4}\Gamma^\gamma_{1\rightarrow0}$. The CMB color temperature at 
the 21-cm energy is estimated in the Reyleigh-Jeans such that $T^{\rm L}_{\rm CMB}=-\frac{E_{21}}{2}$.

\section{BSM Example - Millicharged DM Component}\label{Sec:BSM}
As an example of a new-physics application of \Gcode, we implement an alternative IGM thermal evolution
incorporating the two-component dark matter (2cDM) framework proposed in~\cite{Liu:2019knx}. In this framework, a small fraction of the dark matter energy density, $f_{\rm m}=\Omega_{\rm mDM}/\Omega_{\rm DM}$, consists of fractionally charged (millicharged) particles with mass $m_{\rm m}$ and charge $Q$, which elastically scatter with the baryonic gas and enable heat exchange between the two fluids. For charges large enough to produce observable imprints, CMB constraints imply that the interacting component must remain tightly coupled to baryons prior to recombination and must constitute only a small fraction of the dark matter abundance, $f_{\rm m}\lesssim3\times10^{-4}$~\cite{dePutter:2018xte,Boddy:2018wzy,Barkana:2018qrx}\footnote{This tightly coupled subcomponent mimics the behavior of baryons and its imprints on the CMB are well approximated by a change in the helium fraction, which must remain bellow experimental uncertainties.~\cite{dePutter:2018xte}.}. The remaining dark matter is assumed to consist of cold dark matter (CDM) particles, of mass $m_{\rm C}$, which interact with the millicharged component through a long-range force characterized by a momentum-transfer cross section $\sigma_T^{\rm mC}=\sigma_0 v^{-4}$, thereby effectively increasing the thermal reservoir available to absorb heat from the baryons.

Prior to recombination, the mDM–CDM interaction must remain sufficiently weak to avoid decoupling the millicharged component from baryons or inducing an observable drag on the baryonic fluid~\cite{Boddy:2018wzy,Liu:2019knx}. However, as the relative velocity dissipates toward lower redshifts, the mDM-CDM interaction is enhanced by the $v^{-4}$ velocity dependence, enabling heat transfer between the two dark-sector components. We assume that the impact of the 2cDM model on structure formation, and hence on the HMF, remains subdominant\footnote{While this assumption is certainly valid for a standard fraction of mDM with $f_{m}<3\times10^{-4}$, it would be interesting to check if the late time coupling of CDM to mDM can suppress small scale structures (see~\cite{Driskell:2022pax} for a related study).}. Therefore, in terms of \Gcode\ observables, the 2cDM model results in a reduced baryon temperature during Cosmic Dawn which leads to an enhanced global 21-cm absorption signal, see Sec.~\ref{sub:T21}.

Given the modular structure of \Gcode, implementing the 2cDM model is straight forward as one must only change the IGM evolution module. To account for heat exchange between mDM and baryons, we introduce an additional heat transfer term to the kinetic temperature evolution equation, Eq.~\ref{eq:Tk evolution},
\begin{equation}
    \dot{Q}_{\text{m}\rightarrow \text{b}} =  \sum_{i={\rm e,p,HI,HeI}} \frac{x_i}{1+x_e+f_{He}} \dot{Q}_{{\rm m} \rightarrow i} \, .
    \label{eq:QBSM}
\end{equation}
This additional term sums over the energy transfer rate from the millicharged fluid to each baryonic IGM component (where we neglect interactions with ionized helium for simplicity~\footnote{mDM–HeII interactions become relevant only at late stages of reionization, when the 21-cm signal is already suppressed. }), weighted by its relative abundance in the IGM.

The heat transfer rate between the mDM and each IGM component is given by~\cite{Liu:2019knx}
\begin{equation}\label{eq: Qcool}
\dot{Q}_{{\rm m}\rightarrow i} =  \frac{m_{\rm m} }{m_i+m_{\rm m}} \mu_{i{\rm m}} \langle \Gamma_{i} \vec{v}_{\rm bm}\cdot \vec{V}_{\text{bm}}\rangle+\frac{T_{\rm m}-T_i}{m_i+m_{\rm m}}\frac{\mu_{i{\rm m}}}{u^{i{\rm m}}_{\rm th}}\langle\Gamma_{i} \vec{v}_{\rm bm}\cdot(\vec{v}_{\rm bm}-\vec{V}_{\rm bm})\rangle\ .
\end{equation}
Above, $\mu_{i{\rm m}}$ is the mDM-i reduced mass, and $\Gamma_{i} = n_{\rm m}\sigma_{\rm T}^{i{\rm m}}(v_{\rm bm})v_{\rm bm}$ is the rate at which a single particle of type-$i$ scatters with the entire mDM abundance, where $\sigma_{\rm T}^{i{\rm m}}(v_{\rm bm})$ is the i-mDM momentum transfer cross section as derived in~\cite{Liu:2019knx}, and $\vec{v}_{\rm bm}$ is the relative velocity between the baryonic and mDM fluids.  The angled brackets represent a thermal average where we assume a Gaussian velocity distribution with mean $V_{\rm bm}$ and variance $u_{\rm th}^2=T_i/m_i+T_{\rm m}/m_{\rm m}$. The first term of  Eq.~\eqref{eq: Qcool} accounts for the drag between the two fluids, which dissipates their relative bulk motion into heat. The second term includes the thermal energy transfer, which can either induce heating or cooling, depending on the temperature hierarchy of the two fluids. Once velocities dissipate, the CDM-mDM interactions keep the mDM colder than baryons and 
Eq.~\eqref{eq: Qcool} induces a strong cooling of the baryonic fluid.

To compute the heat transfer rate between baryons and mDM at any redshift, we must self consistently evolve the temperatures of the two DM fluids and the relative  bulk velocities~\cite{Liu:2019knx}
\begin{subequations}\label{eq:All ODES}
    \begin{flalign}
    &\frac{dT_{\rm{m}}}{dz}=\frac{2T_{\rm{m}}}{1+z}-\frac{2}{3k_{\rm B}}\frac{(\dot{Q}_{\rm{mC}}+\dot{Q}_{\rm{mb}})}{H(z)(1+z)}\, ,& \label{eq:evolution macroscopic Tm}\\
    &\frac{dT_{\rm{C}}}{d z}=\frac{2T_{\rm{C}}}{1+z}-\frac{2}{3k_{\rm B}}\frac{\dot{Q}_{\rm{Cm}}}{H(z)(1+z)}\, ,&\label{eq:evolution macroscopic TC}\\
    &\frac{dV_{\rm{bm}}}{dz}=\frac{V_{\rm{bm}}}{1+z}+\frac{\dot{V}_{\rm{bm}}}{H(z)(1+z)}\, ,&\label{eq:evolution macroscopic Vbm}\\
    &\frac{dV_{\rm{mC}}}{dz}=\frac{V_{\rm{mC}}}{1+z} +\frac{\dot{V}_{\rm{mC}}}{H(z)(1+z)}\, ,\label{eq:evolution macroscopic VmC}
    \end{flalign}
\end{subequations}
where the drag between fluids i and j is given by~\cite{Liu:2019knx}
\begin{equation}
\dot{V}_{ij}= \frac{1}{V_{ij}} \frac{m_i\langle  \Gamma_i \vec{v}_{ij}\cdot \vec{V}_{ij}\rangle\ +m_j\langle  \Gamma_j \vec{v}_{ij}\cdot \vec{V}_{ij}\rangle\ }{m_i+m_j}.
\end{equation}
Compton scattering between mDM and CMB photons is not included in Eq.\eqref{eq:evolution macroscopic Tm} since the viable region of the 2cDM sector is in the regime where thermal decoupling of mDM from CMB occurs before recombination~\cite{Liu:2019knx}.

To generate the 2cDM history we initialize \Gcode \ at $z=1500$ when the mDM is still coupled to the baryons. The initial mDM temperature is therefore set to 
$T_{\rm m} = T_{\rm k} = T_\gamma$. We initialize the CDM temperature at an arbitrary small value of $T_{\rm C} = 10^{-3} {\rm K}$ after verifying that the results are largely insensitive to the exact value. Finally, following~\cite{Liu:2019knx} we initialize velocities at $V_{\rm mb}=0$ and $V_{\rm mC}=29\ {\rm km \ s^{-1}}$ and hold them down to $z=1010$. The first is justified by the tight coupling between mDM and baryons, and the second corresponds to the root-mean-square primordial baryon–CDM streaming velocity in $\Lambda$CDM~\cite{Tseliakhovich_2010, Ali_Ha_moud_2014}. 

\begin{figure*}[h!t]
  \centering
  
  \hfill
  \begin{subfigure}[t]{0.48\textwidth}
  \includegraphics[width=\textwidth]{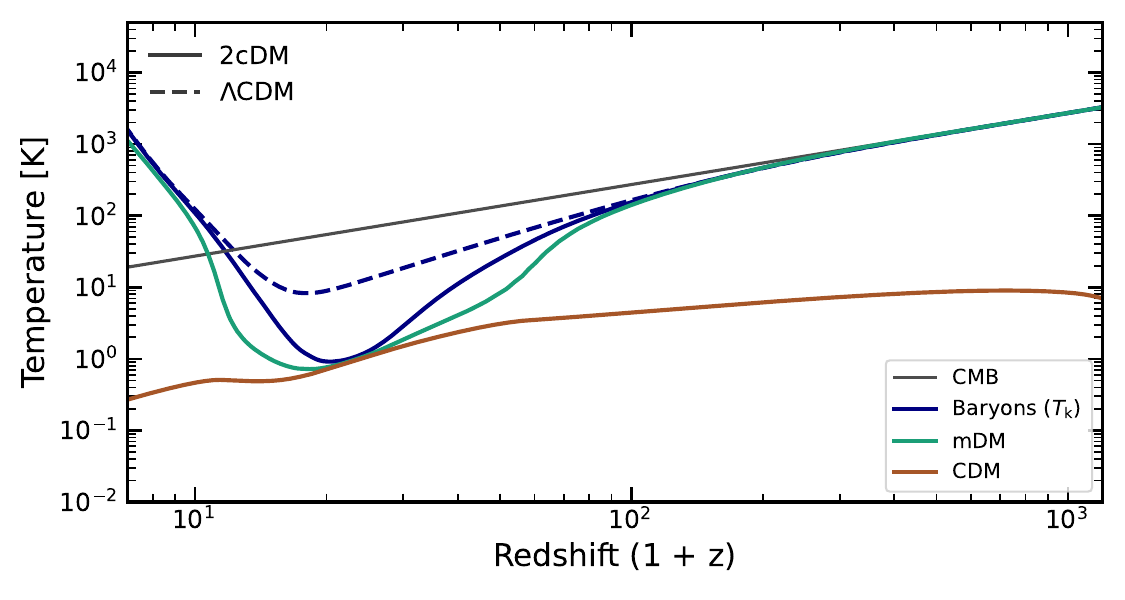}
  \end{subfigure}
  \begin{subfigure}[t]{0.48\textwidth}
    \includegraphics[width=\textwidth]{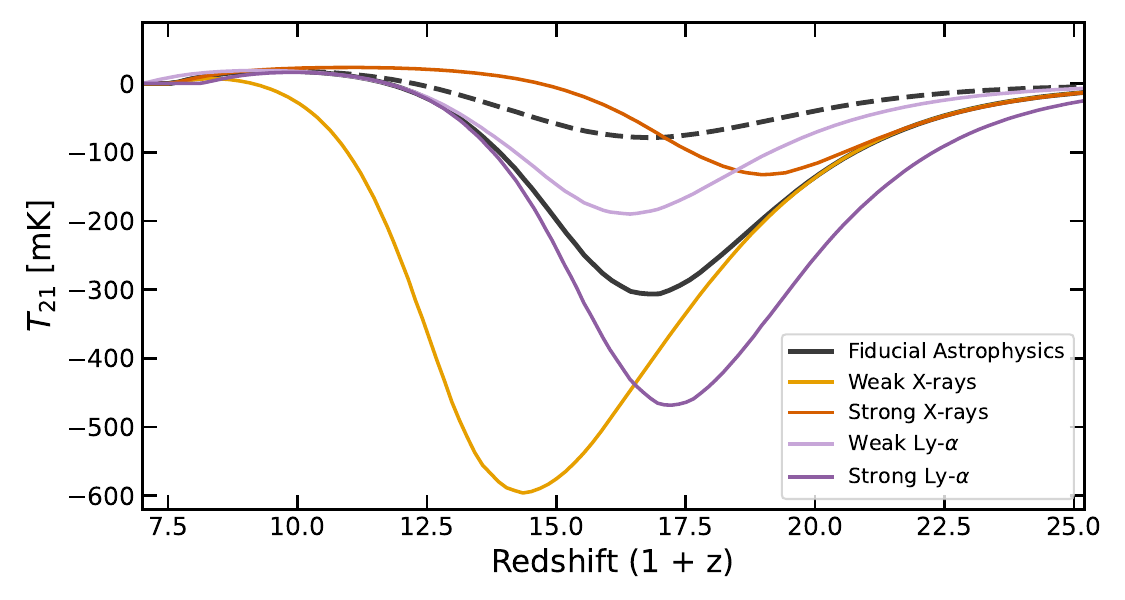}
  \end{subfigure}

  \caption{\textbf{Solid} curves show the thermal and global 21-cm evolution for a viable 2cDM model with $Q=2\times10^{-3}$, $m_{\rm m}=500,{\rm MeV}$, $m_{\rm C}=200,{\rm MeV}$, and $\sigma_0$ set to its maximal value consistent with Planck constraints~\cite{Liu:2019knx}. This parameter choice satisfies the bounds from BBN, accelerator experiments, and direct-detection searches~\cite{Barkana:2018qrx,Munoz:2018pzp,Berlin:2018sjs,Creque-Sarbinowski:2019mcm,Davidson:2000hf,Badertscher:2006fm,Prinz:1998ua,Magill:2018tbb,chatrchyan2013search,ArgoNeuT:2019ckq,Liu:2019knx,Emken:2019tni}, summarized in~\cite{Katz:2024ayw}. The corresponding $\Lambda$CDM evolution is shown with \textbf{dashed} curves.
\textbf{Left:} Temperature evolution of the CMB (\textbf{grey}), baryons (\textbf{blue}), IDM (\textbf{turquoise}), and CDM (\textbf{brown}) assuming the fiducial astrophysical parameters of this work (Tab.~\ref{tab:Params}). Initially, the mDM is tightly coupled to the baryons. After baryon–CMB decoupling at $z\sim250$, the baryon–mDM fluid cools adiabatically until mDM–CDM interactions become efficient at $z\sim100$, leading to a coupled evolution of all three components. This coupling breaks at $z\sim20$, when X-rays by astrophysical sources reheat the baryons, and the mDM with them.
\textbf{Right:} Evolution of the global 21-cm signal. \textbf{Black} curves correspond to the fiducial astrophysical model. The reduced baryonic temperature in the 2cDM scenario produces a deeper absorption feature, enhanced by a factor of $\sim3$ relative to $\Lambda$CDM. \textbf{Purple} curves illustrate the effect of varying the UV normalization by a factor of two around the fiducial value. A larger UV photon production rate increases the \Lya\ background, strengthening the coupling between the spin and kinetic temperatures and resulting in a deeper absorption signal. \textbf{Orange} curves show the impact of varying the X-ray normalization by a factor of five. Stronger (weaker) heating produces a shallower (deeper) absorption trough, as reduced X-ray heating allows $T_{\rm k}$ to remain low at lower redshifts where the \Lya\ intensity is larger, and the spin-kinetic coupling is tighter.}
  \label{fig:T21_DM}
\end{figure*}

In Fig.~\ref{fig:T21_DM} (left) we show the thermal evolution for a viable 2cDM parameter choice, assuming the fiducial astrophysical model of this work. The significant decrease in the baryon kinetic temperature, $T_{\rm k}$, relative to the $\Lambda$CDM scenario is clearly visible. The resulting impact on the global 21-cm signal is shown as a black curve in Fig.~\ref{fig:T21_DM} (right), where the additional cooling produces an enhanced Cosmic Dawn absorption feature compared to $\Lambda$CDM. The 2cDM model also predicts a deeper absorption signal during the Dark Ages, when the spin temperature couples to the baryons through collisions; however, this epoch remains beyond the reach of current experiments~\cite{Liu:2022iyy}.

Large enhancements of the Cosmic Dawn signal can, in principle, be constrained by global 21-cm observations~\cite{Katz:2024ayw}. However, as further illustrated in Fig.~\ref{fig:T21_DM} (right), the interplay between 2cDM and astrophysical parameters plays an important role. For fixed 2cDM parameters, the amplitude of the signal can be significantly suppressed by strong X-ray emission, which counteracts cooling, or by a low UV intensity, which weakens the coupling between the spin and kinetic temperatures. Conversely, the absorption feature can become even stronger if X-ray heating is reduced or if the UV emission is enhanced. 
For this reason, tools such as \Gcode, which enable large parameter-space explorations and self-consistently predict multiple observables that constrain the underlying astrophysical parameters, are essential for breaking degeneracies between astrophysics and new physics and deriving robust constraints on the latter, as demonstrated in~\cite{Katz:2024ayw} for the 2cDM model.

\section{Summary and Discussion}
\label{sec:Summary}
In this work we presented \Gcode, a modular and computationally efficient framework for modeling global observables from Cosmic Dawn and the Epoch of Reionization within a single self-consistent framework. A central motivation for this approach is the increasing availability of complementary probes of the high-redshift Universe, including the global 21-cm signal~\cite{Bowman_2018,Singh:2021mxo}, UV luminosity functions~\cite{Bouwens:2014fua,Bouwens_2021,donnan2024jwstprimernewmultifield,mcleod2023galaxyuvluminosityfunction}, the unresolved cosmic X-ray background~\cite{Hickox_2006,Lehmer:2012ak,Cappelluti:2012rd,Harrison_2016}, and observational constraints on the reionization history~\cite{Planck:2018vyg,McGreer:2014qwa,Davies_2025,Mason_2019}. \Gcode\ is constructed to provide simultaneous predictions for this broad set of observables, allowing their joint interpretation in terms of both astrophysical and new-physics parameters. These predictions are obtained by self-consistently modeling the evolution of high-redshift galaxies and the associated astrophysical radiation backgrounds—including Ly$\alpha$, X-ray, and ionizing radiation fields—together with the thermal and ionization state of the intergalactic medium. The implementation maintains extremely short runtimes of order $\sim 0.1 \ {\rm s}$ per model evaluation, enabling direct parameter inference with modest computational resources (see~\cite{katz2025closingpopiiistarsconstraints} and App.~\ref{Sec:Model}).

A central goal of \Gcode\ is to provide a flexible platform in which astrophysical and BSM extensions can be implemented in a transparent and controlled manner. The modular structure (App.~\ref{Sec:Structure}) allows individual physical components—such as the HMF, star-formation prescriptions, radiation spectra, feedback processes, or additional heat-exchange channels—to be modified independently while preserving a consistent treatment of all aforementioned observables.  
As an example on the BSM side, we showed the straightforward incorporation of an alternative thermal history (Sec.~\ref{Sec:BSM}), including a scenario involving interacting multi-component dark matter, which results in anomalously large 21-cm signals at Cosmic Dawn.

Stochasticity in galaxy properties and star-formation histories has been shown to affect global observables (see e.g.~\cite{Nikoli__2024}). The default astrophysical model adopted in \Gcode\ is currently deterministic; however, such effects can be incorporated straightforwardly within the framework by averaging over parameter distributions. As an example, the implementation of the model by~\cite{Davies_2025_1} was shown in Figs.~\ref{fig:T21} and~\ref{fig:xHI}. A complete treatment including stochastic astrophysical parameters will be explored in future work. Importantly, such extensions preserve the modular structure and computational efficiency of the framework, allowing increasingly sophisticated models to be explored without compromising its suitability for inference studies.

In summary, \Gcode\ provides a fast, flexible, and physically transparent framework for studying the coupled evolution of galaxies, radiation backgrounds, and the intergalactic medium during Cosmic Dawn and the Epoch of Reionization. Its computational efficiency and modular architecture make it particularly well suited for inference studies and for exploring new astrophysical models and extensions beyond the standard model, enabling systematic comparisons between astrophysical and new-physics explanations of upcoming observations.

\acknowledgments
O.Z.K. thanks Tomer Volansky and Diego Redigolo for their continued support and guidance. O.Z.K. also thanks Nadav J. Outmezguine, Hongwan Liu, Andrei Mesinger, and Ely D. Kovetz for many useful discussions.

\bibliography{bib.bib}

\clearpage
\newpage
\appendix

\section{Model Parameter}
\label{Sec:Model}
Our fiducial model, shown throughout this work (e.g., Fig.~\ref{fig:T21}), corresponds to the highest-likelihood solution obtained from an MCMC inference following the procedure of~\cite{katz2025closingpopiiistarsconstraints}. The analysis combines constraints from UVLFs measured by HST~\cite{Bouwens:2014fua,Bouwens_2021}, the unresolved CXB by Chandra~\cite{Cappelluti:2012rd}, the electron scattering optical depth measured by Planck~\cite{Planck:2018vyg}, and constraints on the neutral hydrogen fraction, $x_{\rm HI}(z)$ from quasar absorption spectra~\cite{McGreer:2014qwa,Davies_2025} (see Secs.~\ref{sub:UVLFs},~\ref{sub:CXB} and~\ref{sub:ReionObs} respectively). 

The inference was performed using the public \texttt{emcee} package~\cite{Foreman_Mackey_2013}, requiring a total of $1,125,000$ likelihood evaluations which ran for less than 10 hours on a laptop. The adopted prior ranges and the resulting highest-likelihood parameter values are listed in Tab.~\ref{tab:Params}. We note that the X-ray parameters were not exhaustively explored due to significant degeneracies among them. An investigation of these degeneracies will be presented in forthcoming work~\cite{Xray}. The LW and baryon–DM streaming velocity feedback parameters were chosen to according to simulation results following~\cite{Munoz:2021psm}.

The extended model shown in Fig.~\ref{fig:UVLF} was obtained through a similar inference procedure, with the addition of a free parameter describing the scatter in UV magnitude and by including JWST UVLF measurements in the likelihood. The corresponding best-fit parameters are listed in Tab.~\ref{tab:Params}.

\begin{table*}[t]
\centering
\begin{tabular}{l l l l l}
\hline\hline
\multicolumn{5}{c}{\textbf{Star-formation parameters}} \\
\hline
Parameter & Prior range & HL (Fiducial) & HL (Extended) & Description \\
\hline

$\eps_t$            & $[0,1]$ & 0.38 & 0.84 & Star-formation timescale \\

$\log_{10}(F_\star^{\rm II})$ & $[-3,3]$ & $0.074$ & -0.42 & Pop-II SFE normalization \\

$\log_{10}(M_{\rm p}/M_{\odot})$ & $[8,15]$ & $13.1$ & 13.5 & Pop-II SFE pivot mass \\

$\alpha_{\star}^{\rm II}$ & $[-1,0]$ & $-0.46$ & $-0.31$ & Pop-II low-mass SFE slope \\

$\beta_{\star}^{\rm II}$ & $[0,1]$ & $0.60$ & 0.09 & Pop-II high-mass SFE slope \\

$\log_{10}(M_{ 0}^{\rm II}/M_{\odot})$ & $[6,10]$ & 7.4 & 7.4 & Pop-II cutoff mass \\

$\log_{10}(F_\star^{\rm III})$ & $[-3,1]$ & $-2.3$ & -2.6 & Pop-III SFE normalization \\

$\alpha_{\star}^{\rm III}$ & $[-1,1]$ & $-0.62$ & $0.21$ & Pop-III SFE slope \\

$\log_{10}(M_{0}^{\rm III}/M_{\odot})$ & $[6,\log_{10}(M_{\rm 0}^{\rm II}/M_{\odot})]$ & $6.4$ & 7.2 & Pop-III cutoff mass\\

$A_{\rm LW}$ & -- & $2$ & $2$ & LW feedback normalization \\

$B_{\rm LW}$ & -- & $0.6$ & $0.6$ & LW feedback slope \\

$A_{\rm v}$ & -- & $1$ & $1$ & Velocity feedback normalization \\

$B_{\rm v}$ & -- & $1.8$ & $1.8$ & Velocity feedback slope \\

\hline\hline
\\[-6pt]
\multicolumn{5}{c}{\textbf{X-ray parameters}} \\
\hline
Parameter & Prior range & HL (Fiducial) & HL (Extended) & Description \\
\hline

$\log_{10}\!\left(\frac{\rm (L/SFR)^{II/III}}{\dot{\rm erg \ s^{-1} \ M_{\odot}^{-1} \ yr}}\right)$ 
& [-3,1] & $40.5^{\dagger}$ & $40.5^{\dagger}$ & Pop-II/III X-ray normalization \\

$E_{\min}$ & [0.19,0.85] & $0.5 \ {\rm keV}^\dagger$ & $0.5 \ {\rm keV}^\dagger$ & Minimum escape energy \\

$E_{\max}$ & -- & $30\ {\rm keV}$ & $30\ {\rm keV}$ & High energy cutoff \\

$\alpha_s$ & -- & 1 & 1 & Soft-band slope \\

$\alpha_h$ & -- & 2.2 & 2.2 & Hard-band slope \\

$E_{\rm break}$ & -- & $2 \ {\rm keV}$ & $2 \ {\rm keV}$ & Break energy \\

\hline\hline
\\[-6pt]
\multicolumn{5}{c}{\textbf{Reionization parameters}} \\
\hline
Parameter & Prior range & HL (Fiducial) & HL (Extended) & Description \\
\hline

$\log_{10}(F_{\rm esc}^{\rm II})$  & $[-3,1]$ & $-1.6$ & $-1.3$ & Pop-II escape fraction \\

$\alpha_{\rm esc}^{\rm II}$  & $[-1,1]$ & $0.83$ & 1.2 & Pop-II escape index \\

$N_{\rm \gamma/b}^{\rm II}$  & -- & $5000$ & $5000$ & Pop-II ionizing photons per baryon \\

$\log_{10}(F_{\rm esc}^{\rm III})$ & $[-3,1]$ & $-2.4$ & $-2.9$ & Pop-III escape fraction \\

$\alpha_{\rm esc}^{\rm III}$ & $[-1,1]$ & $0.79$ & $-0.56$ & Pop-III escape index \\

$N_{\rm \gamma/b}^{\rm III}$  & -- & $44021$ & $44021$ & Pop-III ionizing photons per baryon \\

\hline\hline
\end{tabular}
\caption{Parameters of the fiducial and extended astrophysics models adopted in this work, showing the highest-likelihood values from the two inferences, together with the adopted prior ranges and descriptions. Cosmology was set to Planck 2018~\cite{Planck:2018vyg}. $\dagger$ indicates that a different value was inferred; due to degeneracies, we instead adopt the literature value listed here without impacting the likelihood. For X-ray parameters, this is often possible since Chandra provides only upper limits and their impact on reionization is typically negligible.}
\label{tab:Params}
\end{table*}

\begin{figure*}[t]
    \centering
    \includegraphics[width=0.7\textwidth]{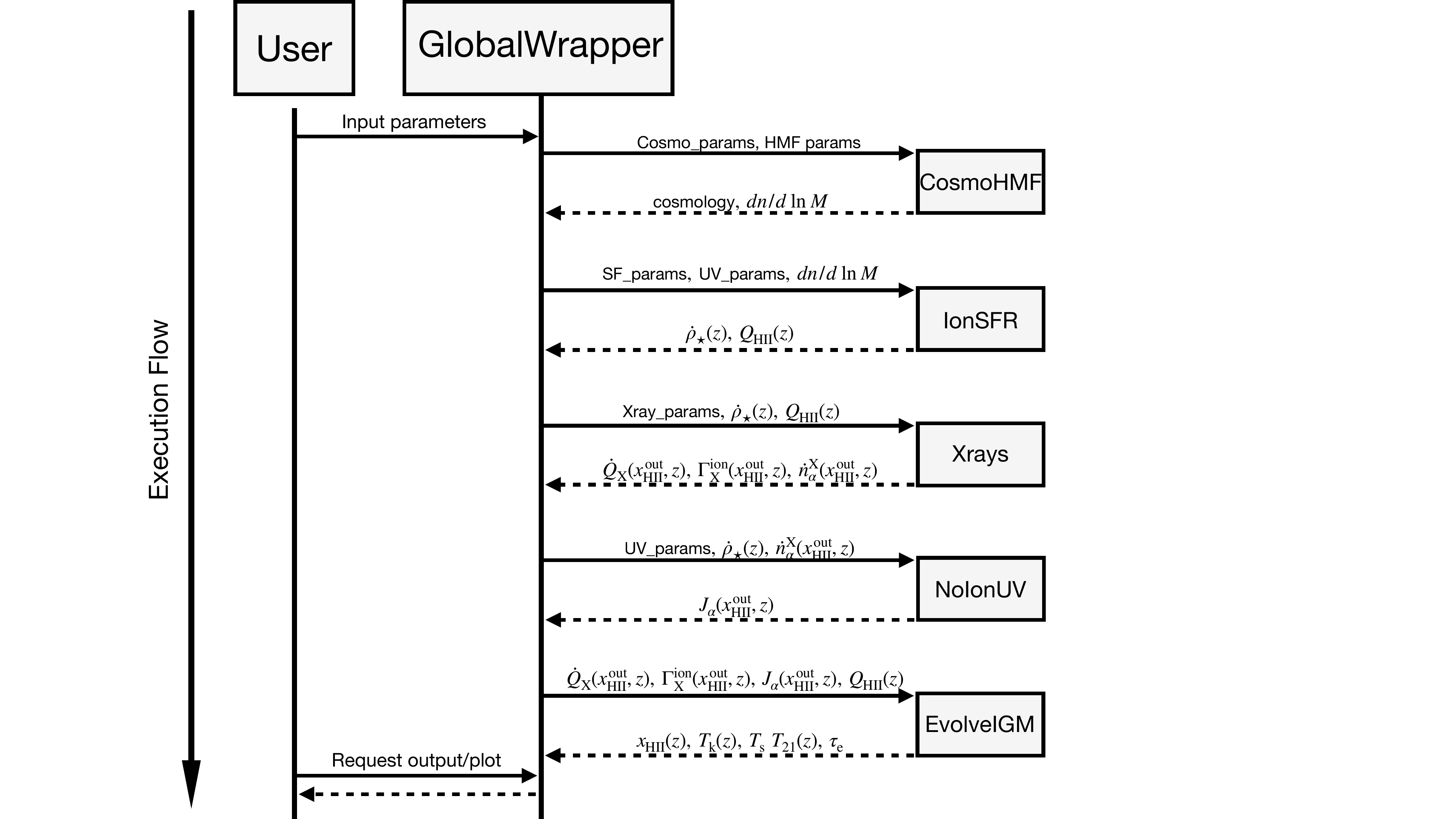}
    \caption{
    Execution flow in \Gcode. Time ordering proceeds from top to bottom.
    The user interacts only with the \texttt{GlobalWrapper} object, which
    sets the cosmology, initializes the astrophysical history and subsequently runs the IGM
    evolution. During initialization, the \texttt{CosmoHMF} module defines the cosmological background and computes the HMF. The HMF is then passed to the \texttt{IonSFR} module, which determines the coupled evolution of the SFRD, UV ionized fraction, and LW photons (see Sec.~\ref{SubSec:SFR}). These outputs are supplied to the \texttt{Xrays} module to compute X-ray heating, ionization rates, and  \Lya\ emissivity from secondary events. The X-ray–induced \Lya\ emissivity is then passed to the \texttt{NoIonUV} object, which computes the direct stellar \Lya\ intensity and combines both contributions to obtain the total \Lya\ background. The resulting radiation backgrounds and ionization histories are then passed to the \texttt{EvolveIGM} module, which solves the coupled evolution equations for the IGM thermal and ionization state. This module also computes the 21-cm brightness temperature and the hydrogen spin temperature, the latter entering the thermal evolution when \Lya\ heating is included (see Sec.~\ref{SubSec:Tk}). Overall, a full execution of the pipeline typically requires $\sim 0.1 \  \mathrm{s}$. For flexability, the CXB and UVLFs are computed post-evolution on demand via the \texttt{GlobalWrapper}, which internally calls the \texttt{Xrays} and \texttt{IonSFR} modules.}
    \label{fig:sequence}
\end{figure*}

\section{Code Structure}\label{Sec:Structure}

The main goal of \Gcode\ is to relate astrophysical and, when desired, new-physics models to observables from CD. Its short runtime enables efficient parametric inference over large multidimensional parameter spaces, as demonstrated in~\cite{katz2025closingpopiiistarsconstraints,Katz:2024ayw}. To facilitate model exploration, \Gcode\ is designed with a simple and modular structure, illustrated in Fig.~\ref{fig:sequence}.

A full execution typically requires $\sim 0.1 \ \mathrm{s}$, after which the user may query a range of astrophysical and cosmological outputs. The main outputs include the global average SFRD, UVLFs at user-specified redshifts, the contribution of X-ray sources to the present-day CXB within a user-specified energy band, ionization history and the optical depth to reionization, the global 21-cm signal, and hydrogen spin and kinetic temperatures. All computed astrophysical and IGM quantities are accessible directly through the \texttt{GlobalWrapper} object which also offers a built in plotter.

As an example of a new-physics application within \Gcode, we provide an alternative IGM evolution module, \texttt{EvolveIGM\_mDM}, which replaces the standard \texttt{EvolveIGM}. This module evolves both the standard IGM components and two dark matter components following~\cite{Liu:2019knx} as described in Sec.~\ref{Sec:BSM}.

\end{document}